\documentclass[aps,prfluids,superscriptaddress,showpacs,amsmath,amssymb,nofootinbib,reprint]{revtex4-2}
\usepackage{amsfonts}
\usepackage{amsthm}
\usepackage{graphicx}
\usepackage{mathtools}
\usepackage{color}
\usepackage{url}
\usepackage[abs]{overpic}
\usepackage[usenames,dvipsnames]{xcolor}
\usepackage[normalem]{ulem}
\usepackage{tabularx}

\def\kmax{k_{\rm max}}

\newcommand\Rey{{\text{Re}}}

\newcommand{\ex}{{\bf e}_x}
\newcommand{\ez}{{\bf e}_z}

\newcommand\Rb{R_b }
\newcommand\Fr{\text{Fr}}

\def\mbfx{\mathbf{x}}

\def\mbfu{\mathbf{u}}
\def\mbfk{\mathbf{k}}

\begin{document}

\title{Energy transfer, Intermittency and Mixing in Shear-Driven Stratified Turbulence}

\author{Chandra Shekhar Lohani}
\email{cslohani25@gmail.com}
\affiliation{Department of Physics, Indian Institute of Technology Kharagpur, Kharagpur - 721 302, India}
\author{Vishwanath Shukla}
\email{vishwanath.shukla@phy.iitkgp.ac.in}
\affiliation{Department of Physics, Indian Institute of Technology Kharagpur, Kharagpur - 721 302, India}%

\date{\today}
\begin{abstract} 
We investigate a stably stratified flow driven by deterministic Kolmogorov forcing that generates horizontal shear, using direct numerical simulations over a broad range of stratification strengths characterized by the Froude number $\Fr$. As the stratification is progressively weakened, the flow exhibits a sequence of regimes: a buoyancy-dominated, strongly stratified regime, an intermediate regime characterized by Kelvin--Helmholtz instabilities and enhanced mixing, and a nearly isotropic turbulent regime. A key feature of the intermediate stratification range is the emergence of energetically significant vertically sheared horizontal flows (VSHFs), accompanied by a marked steepening of the reduced one-dimensional perpendicular kinetic energy spectra. The spectral energy transfer remains predominantly forward, although the perpendicular flux becomes negative at large horizontal scales; this apparent upscale transfer reflects anisotropic energy redistribution rather than a true inverse cascade. Strong stratification enhances intermittency, producing increasingly non-Gaussian vertical velocity fluctuations and large kurtosis associated with localized vertical bursts. The energetics-based mixing coefficient remains of order $10^{-1}$ over the parameter range investigated, with a modest enhancement near the Kelvin--Helmholtz instability regime.
\end{abstract}

\maketitle

\onecolumngrid

\section{Introduction}

Stratification strongly influences the evolution of turbulence in geophysical flows and in the interiors of planets and stars, where gradients in temperature and salinity (or analogous scalar fields) give rise to density stratification, introducing new energy pathways and fundamentally altering flow organization~\cite{davidson2013turbulence,verma2018physics}. Stratified flows are known to exhibit quasi-two-dimensional (quasi-2D) structures, often referred to as quasi-horizontal layers or ``pancakes'', which may arise from the destabilization of coherent columnar vortices via the zigzag instability~\cite{billant2000experimental}. The formation of these structures is associated with the suppression of vertical motions by buoyancy forces; consequently, motions along the direction of the imposed stratification are inhibited, leading to significant modifications of the energy cascade and mixing properties in strongly anisotropic regimes.

Based on measurements of atmospheric wind and temperature spectra, an inverse cascade of kinetic energy was initially proposed for purely stratified turbulence~\cite{lilly1983stratified}. Subsequent numerical studies and analyses of atmospheric data, however, demonstrated that energy transfer is predominantly forward, with a horizontal kinetic energy spectrum scaling as $k_{\perp}^{-5/3}$~\cite{lindborg2006energy,brethouwer2007scaling,riley2008stratified,lindborg2008stratified,almalkie2012kinetic}, where $k_{\perp}$ denotes the horizontal wave number. The nature of spectral transport depends sensitively on the stratification strength.

In moderately stratified turbulent flows, buoyancy forces convert kinetic energy into potential energy, leading to the Bolgiano–Obukhov phenomenology. In this buoyancy-dominated regime, the kinetic energy spectrum exhibits a $k^{-11/5}$ scaling at large length scales and transitions to the classical Kolmogorov $k^{-5/3}$ scaling at smaller scales. For weaker buoyancy, the flow approaches a regime in which buoyancy behaves as a passive scalar. In contrast, turbulence under strong buoyancy exhibits scaling behaviour and dynamical characteristics that remain far from fully understood~\cite{obukhov1959influence,bolgiano1959turbulent,kumar2014energy,rosenberg2015evidence,alexakis2018cascades}.

Several approaches have been used to study turbulent flows under strong stable stratification. Wave–vortical decomposition has been used to elucidate the role of anisotropy in energy transport under random, large-scale forcing applied isotropically to the horizontal velocity field~\cite{riley1981direct,kimura2012energy}. In such studies, the vortical component was found to exhibit $k_{\perp}^{-3}$ and $k_{\perp}^{-5/3}$ scalings at small and large wave numbers, respectively~\cite{kimura2012energy}, with the latter range shifting toward higher wave numbers as the stratification strength increases. The wave component displayed $k_{\perp}^{-2}$ scaling at large scales and $k_{\perp}^{-5/3}$ scaling at smaller scales. The corresponding vertical wave number spectrum exhibited a $k_z^{-3}$ behavior at small scales, while remaining flat at large scales.

Linear eigenmode decompositions of the Boussinesq equations have also been used to examine wave and vortical contributions in stratified flows, both with and without rotation~\cite{bartello1995geostrophic,smith2002generation,waite2006transition,herbert2016waves}.
In~\cite{brethouwer2007scaling}, the buoyancy Reynolds number was employed to classify strongly stratified flows driven by purely horizontal two-dimensional (2D) forcing. At low buoyancy Reynolds number, both the horizontal and vertical energy spectra were found to be steep, whereas at large buoyancy Reynolds number the kinetic and potential energy spectra exhibited $k_{\perp}^{-5/3}$ scaling, indicative of a forward energy cascade. The role of buoyancy-scale dynamics in shaping the kinetic energy spectrum was further examined in~\cite{waite2011stratified}. Another key length scale is the Ozmidov scale, $\ell_{oz}$, below which isotropy is recovered. High-resolution simulations in~\cite{almalkie2012kinetic} showed that the horizontal energy spectra exhibit $k_{\perp}^{-5/3}$ scaling both below the Ozmidov scale and in the range between the buoyancy and Ozmidov scales.

The dynamics of stratified flows are governed by the interplay between slow, quasi-2D modes with $k_z \approx 0$ and fast, wave-dominated modes with finite $k_z$, associated with internal gravity waves. Under strong stratification, energy can accumulate spectrally at $k_{\perp} = 0$ and $k_z \neq 0$, giving rise to vertically sheared horizontal flows (VSHFs), also referred to as shear modes~\cite{smith2002generation,laval2003forced,lindborg2006energy,brethouwer2007scaling}. Despite extensive study, the dynamics and formation mechanisms of VSHFs remain incompletely understood~\cite{fitzgerald2018statistical}. Resonant interactions among internal gravity wave modes have been identified as a possible formation mechanism~\cite{smith2002generation}, while secondary instabilities have also been shown to play a key role in their development~\cite{fitzgerald2018vertically}.

An additional key feature of stably stratified flows is the presence of strong intermittency in the vertical velocity and density (or temperature) fields, associated with the spontaneous development of long-lasting bursts. Such intermittency leads to extreme events and is manifested by the emergence of non-Gaussian tails in the probability distribution functions (PDFs) of these fields, with stronger stratification giving rise to enhanced intermittency. In~\cite{feraco2018vertical}, both Lagrangian and Eulerian data were used to link the non-Gaussian tails of the PDFs to large-scale bursts generated by the interplay between internal gravity waves and turbulent motions across a range of stratification strengths relevant to geophysical flows.  

Intermittency has also been associated with local overturning events and the resulting mixing, which may be characterized by the gradient Richardson number, mixing efficiency, and the ratio of kinetic to potential energy. These mixing processes have important implications for phytoplankton blooms in the ocean and the dispersion of atmospheric pollutants, both of which occur in naturally stratified environments. The dependence of mixing efficiency on various control parameters, such as the Richardson number, has been examined in a variety of laboratory experiments~\cite{turner1968influence,park1994turbulent,olsthoorn2015vortex}.

Earlier studies have primarily examined stably stratified flows subjected to isotropic three-dimensional (3D) or 2D random forcing, with or without a small vertical forcing component. In contrast, relatively few studies have considered horizontally sheared external forcing. Horizontal shear flows, often supported by zonal jets, are commonly observed in the oceans~\cite{munk2000spirals}. Moreover, decaying horizontally sheared flows have been shown to exhibit more intense turbulence than vertically sheared flows~\cite{caulfield2000anatomy,basak2006dynamics}. More recently, stratified turbulence driven by horizontal shear has been investigated to elucidate the organization of density-field structures and the associated instability and bifurcation mechanisms~\cite{lucas2017layer}.

In this study, we investigate a horizontally sheared, stably stratified flow driven by deterministic horizontal Kolmogorov forcing. The computational domain has a horizontal extent four times its vertical height. This choice reduces computational cost relative to a cubic domain, while remaining sufficiently thick to avoid strong confinement effects and support nearly isotropic dynamics in the absence of stratification. We further choose the ratio of the domain height to the forcing length scale to exceed a critical threshold, below which confinement-induced effects lead to a pronounced inverse energy cascade. We analyze the statistical properties of the flow over a broad range of stratification strengths, from weak to strong, focusing on the evolution of flow structures, suppression of vertical velocity fluctuations, partitioning of energy among wave, eddy, and large-scale mean flows, spectral energy transport, intermittency, and mixing characteristics. Together, these results show how deterministic horizontal shear, stable stratification, and VSHF formation reorganize spectral transport, intermittency, and mixing in stratified turbulence.

\section{Theoretical Framework}

\subsection{Governing equations and parameters}

We consider an incompressible, stably stratified fluid in a triply periodic domain $\mathcal{V}$ of dimensions $L\times L\times H$, with $H=L/4$. The incompressible Navier–Stokes equations under the Boussinesq approximation govern the evolution of the velocity field $\mbfu(\mbfx,t)$ and the scalar fluctuation field $\theta(\mbfx,t)$. The governing equations read 
\begin{subequations}\label{eq:NSEBSQ}
\begin{align}
	\dfrac{\partial {\bf u}}{\partial t} + (\bf{u} \cdot {\bf \nabla}){\bf u} & = -\frac{1}{\rho_0} {\bf \nabla} p + \nu \nabla^2 {\bf u} - N\theta {\bf e}_z+ {\bf f}, \label{eqn:NS} \\
	\dfrac{\partial \theta}{\partial t} + (\bf{u} \cdot {\bf \nabla})\theta & = \kappa \nabla^2 {\theta} + N w, \label{eqn:denfluct} \\
	{\bf \nabla} \cdot {\bf{u}} & = 0, \label{eqn:incompressibility}
\end{align}
\end{subequations}
where $\mbfu=(u,v,w)$, $\rho_0=1$ is the fluid density, $p(\mbfx,t)$ is the pressure, $\nu$ is the kinematic viscosity, $\kappa$ is the diffusivity, and $N$ is the Brunt--V\"ais\"al\"a frequency. Here $\theta(\mbfx,t)$ denotes a buoyancy (or temperature) fluctuation field, linearly related to density and temperature fluctuations under the Boussinesq approximation and rescaled to have units of velocity. 

Turbulence is sustained by externally driving the system with a 2D Kolmogorov forcing, $\mathbf{f}=f_0 \cos(k_f y) \,\ex$, where $f_0$ is the forcing amplitude and $k_f$ is the forcing wave number. The forcing is invariant along the $\ez$-direction and introduces shear into the stratified flow~\cite{basak2006dynamics,lucas2017layer}.

To characterize stratified turbulence, we use the Reynolds number $\Rey=UL/\nu$ and the Froude number $\Fr=U/(NL)$, where $U$ denotes the root-mean-square velocity. The Froude number measures the relative importance of buoyancy forces compared to inertial effects and may equivalently be interpreted as the ratio of the gravity-wave period, $N^{-1}$, to the eddy turnover time, $L/U$. Stratified turbulence is characterized by the formation of layered structures, which render the flow anisotropic. The associated buoyancy (vertical) length scale is $\ell_b=U/N$, and shear between these layers gives rise to strong vertical gradients.

At sufficiently small scales, however, flow isotropy is restored below the Ozmidov length scale, $\ell_{oz}=\sqrt{\epsilon_{\mathrm{dis}}^{K}/N^3}$, where $\epsilon_{\mathrm{dis}}^{K}=\nu \langle (\nabla \mbfu)^2 \rangle$ denotes the kinetic energy dissipation rate. The buoyancy Reynolds number, $R_b=\Rey \Fr^2$, measures the ability of turbulent motions to overcome viscous effects in the presence of stratification and thereby quantifies the degree of scale separation between the Ozmidov scale $\ell_{oz}$ and the Kolmogorov scale $\eta$.

The statistical properties of the flow are influenced not only by stratification but also by the domain geometry. The flow domain is compressed in the vertical direction, with an aspect ratio $\gamma=H/L=1/4$. Taking $L$ and $U$ as the characteristic horizontal length and velocity scales, incompressibility implies $w/(\gamma L)\sim U/L$, leading to the scaling $w\sim\gamma U$. As a result, vertical motions are suppressed and the associated kinetic energy scales as $\gamma^2 U^2$.

In addition, spectral transport in confined domains may exhibit the coexistence of forward and inverse energy cascades, with the latter depending sensitively on the aspect ratio. For the present geometry, however, this effect is negligible.

The inclusion of stable stratification further enriches the dynamics of the compactified flow. For $\Fr \ll 1$, the flow is strongly stratified, with buoyancy forces dominating the vertical nonlinear inertial terms. The characteristic vertical (buoyancy) length scale is $\ell_b \sim U/N$, which, together with incompressibility, implies a vertical velocity scale $w \sim \ell_b U/L \approx U^2/(NL)$. Consequently, the kinetic energy associated with vertical motions scales as $E_z/E_K \sim w^2/E_K \sim \Fr^2$, where $E_K \sim U^2$ denotes the total kinetic energy. In this regime, the strongly anisotropic flow organises into layered structures at scales larger than the Ozmidov scale, $\ell \gg \ell_{oz}$.

\subsection{Numerical simulations}

\begin{table*}%[h!]
	%\large
	\begin{center}
		\begin{tabular}{|l||c|c|c|c|c|c|c|c|c|c|c|c|c|}
			\hline\hline
			\texttt{} & \texttt{R0} & \texttt{R1} & \texttt{R2} & \texttt{R3} & \texttt{R4} & \texttt{R5} & \texttt{R6} & \texttt{R7} & \texttt{R8} & \texttt{R9} & \texttt{R10} & \texttt{R11} & \texttt{R12}\\
			\hline
			$N$	   & 0     & 0.5	& 1    & 1.5	& 2     & 3     & 4     & 5     & 6     & 10    & 12    & 20    & 30 \\
			$\Rey$   & 3030  & 3023	& 2746  & 2608	& 2661  & 3311  & 4032  & 3195  & 3141  & 3547  & 4014  & 3000  & 2857 \\
			$\Fr$   & $-$   & 0.224	& 0.104 & 0.066	& 0.051 & 0.042 & 0.038 & 0.024 & 0.020 & 0.013 & 0.012 & 0.006 & 0.004 \\
			$\Rb$  & $-$   & 151.7	& 29.7  & 11.4	& 6.92  & 5.84  & 5.8   & 1.84  & 1.26  & 0.6   & 0.58  & 0.11  & 0.037 \\
			\hline

       $\ell_f/\eta$  & 48 .3   & 49	& 50.4    & 53.9	& 60.3    & 66.9  & 67.3  & 59    &  58   & 60.3    &  64.9   & 61.8    & 59.4 \\
    $\ell_b/\eta$  & $-$   & 87.8 	& 42.3   & 28.5		& 24.4   & 22.2   & 21   & 11.3    &  9.1   & 7.3    &  6.5   &  2.8   & 1.7  \\
 $\ell_{oz}/\eta$  & $-$   & 39.8 	& 15.3    & 10.2	& 9.3    &  6.9   & 4.6     &  2.2   &  1.6   & 0.83   &   0.79   &  0.32    & 0.15   \\
$\ell_b/\ell_{oz}$ & $-$   & 2.21	& 2.8   & 2.79		& 2.6   &  3.2  & 4.6   &  5.2  &  5.9  & 8.7   &  8.2  &  8.8  & 11.2  \\

			\hline
		\end{tabular}
	\end{center}
	\caption{\small 
		Summary of the Direct numerical simulations (DNS) runs.
		$N$ is the Brunt–Väisälä frequency, $\Rey = UL/\nu$ is the Reynolds number, $\Fr$ is the Froude number, and $\Rb$ is the buoyancy Reynolds number. The forcing length scale is $\ell_f$, $\eta$ denotes the Kolmogorov dissipation scale, $\ell_b$ is the buoyancy length scale, and $\ell_{oz}$ is the Ozmidov length scale. The computational domain $\mathcal{V}$ has dimensions $L \times L \times H$, with $L=2\pi$ and $H=\pi/2$. Accordingly, $\ell_f=\pi/4$, corresponding to a forcing wave number $k_f=8$. The simulations use $512^2 \times 128$ collocation points with isotropic grid spacing. The Prandtl number is fixed at $\Pr=\nu/\kappa=1$, and the Reynolds number is maintained nearly constant at $\Rey \sim \mathcal{O}(10^3)$. DNS runs are performed using the GHOST code~\cite{mininni2011hybrid}.
	}
	\label{tab:runs}
\end{table*}

We use a pseudo-spectral method to numerically solve the governing equations Eqs.~\eqref{eq:NSEBSQ}(a)–(c), with aliasing errors removed using the $2/3$ de-aliasing rule. The computational domain is defined by $L=2\pi$ and $H=\pi/2$. To ensure isotropic grid spacing, we use $N_x=512$, $N_y=512$, and $N_z=128$ collocation points along the $x$-, $y$-, and $z$-directions, respectively. The forcing length scale is set to $\ell_f=\pi/4$, corresponding to a forcing wave number $k_f=8$.

The choice of $H>\ell_f$ allows forcing to occur over a 3D wave number range, with the smallest accessible vertical wave number given by $k_z=4$. We note that, in rotating turbulence, the emergence of two-dimensionality in the limit of large rotation rates is independent of whether $H>\ell_f$ or $H<\ell_f$~\cite{smith1996crossover}. Direct numerical simulations are performed using the GHOST code~\cite{mininni2011hybrid}, and the simulation parameters are summarized in Table~\ref{tab:runs}. The Prandtl number is fixed at $\mathrm{Pr}=\nu/\kappa=1$. The Kolmogorov dissipation length scale $\eta$ is well resolved in all simulations, with $\kmax^{\perp}\eta \approx 2.0$, where $\kmax^{\perp}=N_x/3$ denotes the maximum perpendicular wave number. Consequently, the Reynolds number remains moderate, $\Rey \sim \mathcal{O}(10^3)$.

In our DNS runs, the ratio $L/\ell_f=8$ is kept fixed. The scale separation $\ell_f/\eta$ is approximately $60$ for $\Fr \leq \Fr_c$ and decreases to about $50$ for $\Fr > \Fr_c$, where $\Fr_c=0.051$. The separation between the buoyancy length scale and the Kolmogorov scale reduces to less than one decade for $\Fr \lesssim 0.02$, at which point the Ozmidov scale $\ell_{oz}$ enters the dissipation range.

The instantaneous kinetic and potential energies are defined as $E_K(t)=\langle \mbfu^2 \rangle_{\mbfx}$ and $E_P(t)=\langle \theta^2 \rangle_{\mbfx}$, respectively, where $\langle \cdot \rangle_{\mbfx}$ denotes spatial averaging. The kinetic energy associated with the vertical velocity component is given by $E_z(t)=\langle w^2 \rangle_{\mbfx}$. To characterize spectral energy transport in this inherently anisotropic system, we compute the 2D axisymmetric kinetic energy spectrum
\begin{equation}
	E(k_{\perp}, k_{\parallel}) = \frac{1}{2}
	\sum_{\substack{
			k_{\perp} \leq |\mathbf{k} \times \mathbf{e}_z| < k_{\perp} + 1 \\
			k_{\parallel} \leq |\mathbf{k} \cdot \mathbf{e}_z| < k_{\parallel} + 1
	}}
	\left| \widehat{\mathbf{u}}(\mathbf{k}) \right|^2,
\end{equation}
where $k_{\perp}=\sqrt{k_x^2+k_y^2}$ and $k_{\parallel}=|k_z|$. Here $\widehat{\mathbf{u}}(\mathbf{k})$ denotes the Fourier transform of the velocity field $\mbfu$. The reduced one-dimensional (1D) spectra $E(k_{\perp})$ and $E(k_{\parallel})$ are obtained by integrating $E(k_{\perp},k_{\parallel})$ over $k_{\parallel}$ and $k_{\perp}$, respectively. We define the total axisymmetric kinetic energy flux as
\begin{equation}
	\Pi(k_\perp, k_\parallel) = - \sum_{\substack{|\mathbf{k}_\perp| \leq k_\perp \\ |k_z| \leq k_\parallel}} \left[ \widehat{\mathbf{u}}^*(\mathbf{k}) \cdot \widehat{(\mathbf{u} \cdot \nabla \mathbf{u})}(\mathbf{k}) \right],
\end{equation}
from which the reduced fluxes $\Pi(k_{\perp})$ and $\Pi(k_{\parallel})$ along the perpendicular and parallel directions, respectively, are obtained.

We also examine the partitioning of energy among the wave-like (poloidal) and vortical (toroidal) components. Given the incompressible nature of the flow, we introduce the orthonormal Craya--Herring basis~\cite{chandrasekhar2013hydrodynamic,riley1981direct,praud2005decaying,kimura2012energy}, defined by $\mathbf{e}_1=\mathbf{k}\times\mathbf{e}_z/k_{\perp}$, $\mathbf{e}_2=\widehat{\mathbf{k}}\times\mathbf{e}_1$, and $\mathbf{e}_3=\widehat{\mathbf{k}}=\mbfk/|\mbfk|$. The velocity field can then be expressed in this basis, including the VSHF modes with $k_{\perp}=0$, as
\begin{equation}
	\widehat{\mbfu}(\mbfk)=
	\begin{cases}
		& \widehat{u}_v(\mbfk)\mathbf{e}_1 + \widehat{u}_w(\mbfk) \mathbf{e}_2 \quad \quad k_{\perp}\neq 0 \\
		& \widehat{\mbfu}_s \qquad \qquad \qquad \qquad \quad k_{\perp} = 0.
	\end{cases}
\end{equation}
Here $\widehat{u}_v$ and $\widehat{u}_w$ denote the vortical and wave-like components, respectively, while $\widehat{\mbfu}_s$ denotes the vertically sheared horizontal velocity field. The total energy is decomposed as $E=E_v+E_w+E_s$, with $E_v=\langle \widehat{u}_v^2\rangle_\mbfk$, $E_w=\langle \widehat{u}_w^2\rangle_\mbfk+E_P$, and $E_s=\langle \widehat{u}_s^2\rangle_\mbfk$. Here $\langle \cdot \rangle_\mbfk$ denotes averaging over Fourier modes. Thus, the potential energy is included in the wave-like contribution.

%\subsection{Flow regimes}

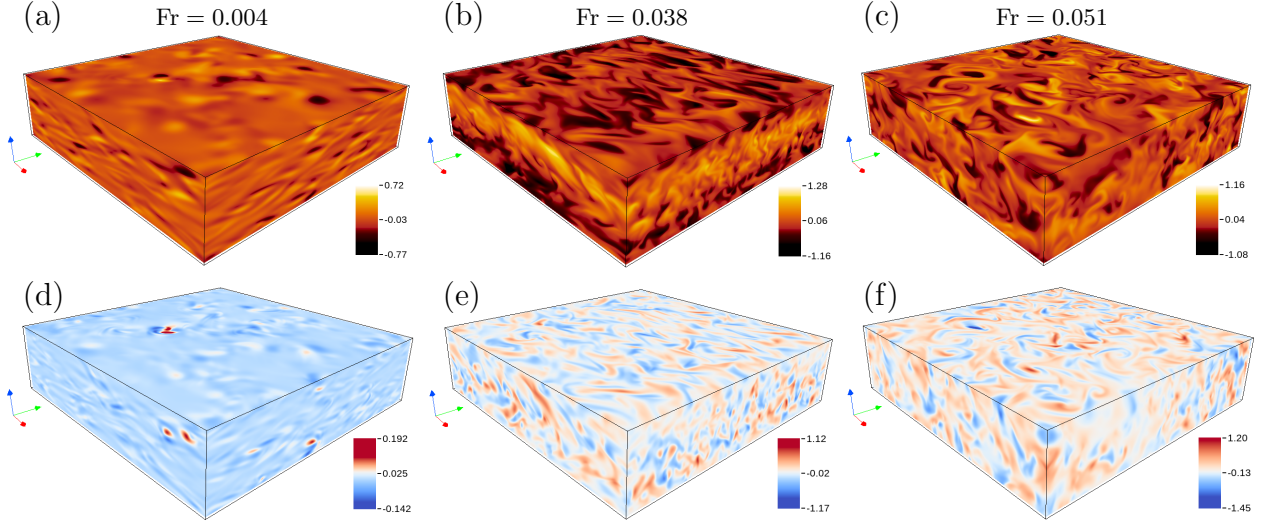
\begin{figure*}
	\centering
	\includegraphics[width=0.31\linewidth]{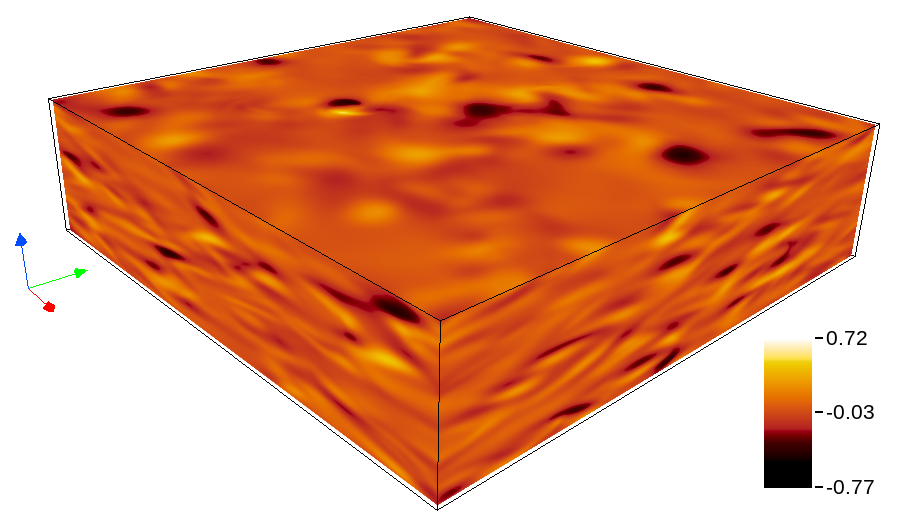}
	\put(-150,94){{\large{(a)}}} \put(-100,94){{$\Fr=0.004$}}
	\includegraphics[width=0.31\linewidth]{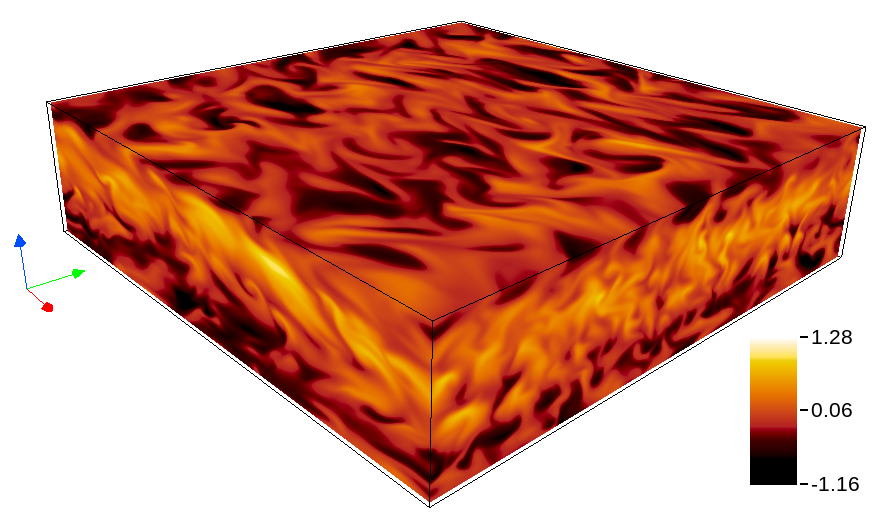}
	\put(-150,94){{\large{(b)}}} \put(-100,94){{$\Fr=0.038$}}
	\includegraphics[width=0.31\linewidth]{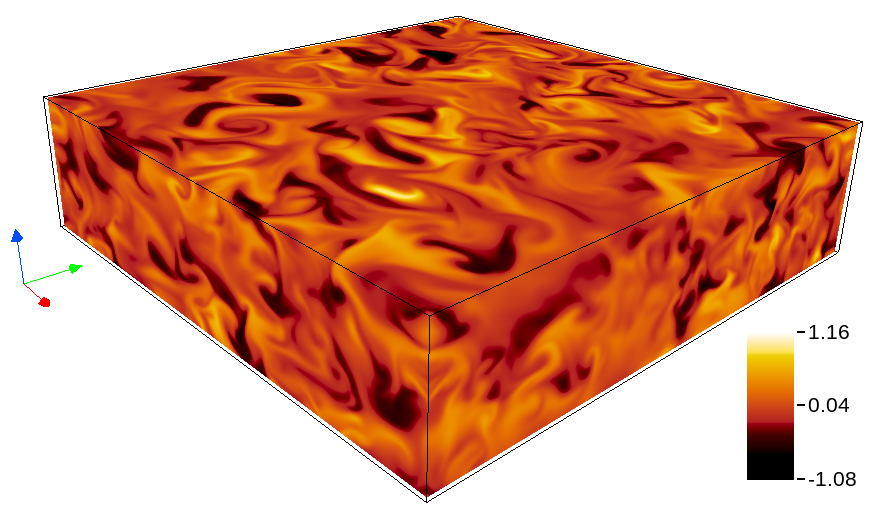}	
		\put(-150,94){{\large{(c)}}} \put(-100,94){{$\Fr=0.051$}}\\
	\includegraphics[width=0.31\linewidth]{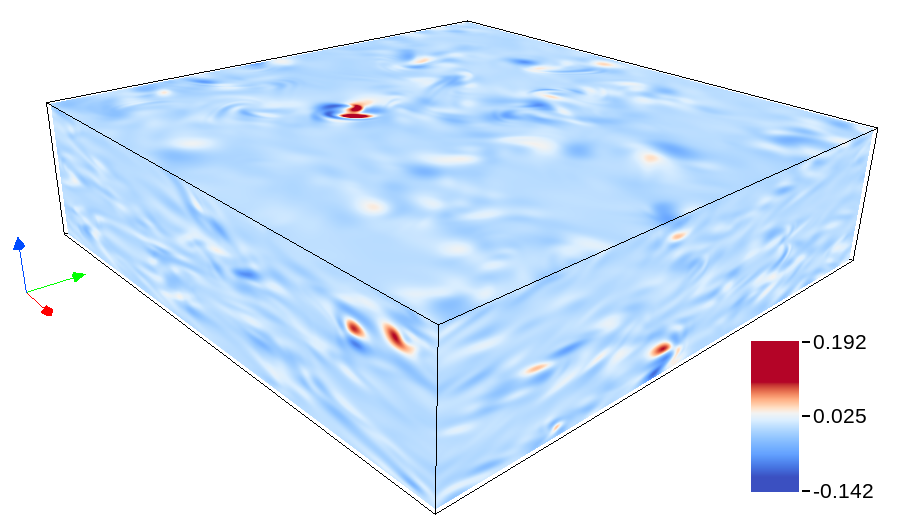}
	\put(-150,84){{\large{(d)}}}
	\includegraphics[width=0.31\linewidth]{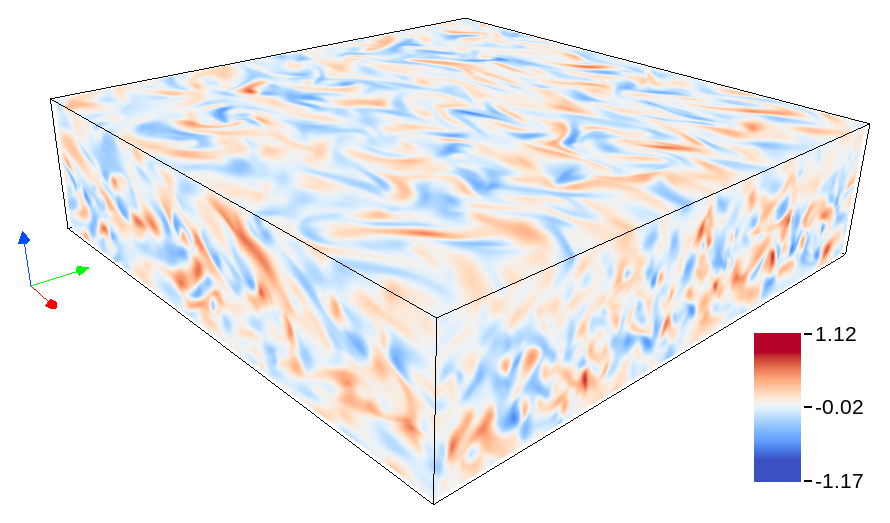}
	\put(-150,84){{\large{(e)}}}
	\includegraphics[width=0.31\linewidth]{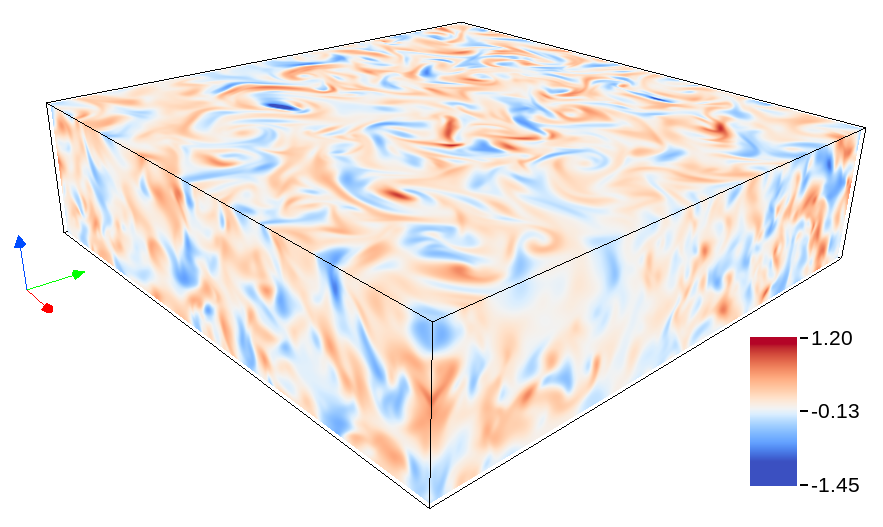}
	\put(-150,84){{\large{(f)}}}
	\caption{ Flow-field visualizations under varying stratification.
		Pseudocolor visualizations of the buoyancy fluctuation field $\theta(\mbfx)$ (upper panels) and the vertical velocity component $w(\mbfx)$ (lower panels) for different Froude numbers: $\Fr=0.004$ (a,d), $\Fr=0.038$ (b,e), and $\Fr=0.051$ (c,f). At the smallest Froude number, $\Fr=0.004$, the $\theta$ field exhibits pronounced V-shaped layering along the vertical direction. At the intermediate value $\Fr=0.038$, overturning regions appear and Kelvin–Helmholtz instabilities develop. The red, green, and blue arrows indicate the $x$-, $y$-, and $z$-directions, respectively. Flow fields are visualized using the VAPOR software.}
	\label{fig:flowfields}
\end{figure*}

\section{Results}

\subsection{Flow organization}

We first describe the spatial structures that develop in the flow by varying the stratification strength while keeping the Reynolds number nearly fixed. Figures~\ref{fig:flowfields}(a)--(c) show pseudocolor plots of $\theta(\mathbf{x},t)$ for $\Fr=0.004$, $0.038$, and $0.051$, respectively; Figures~\ref{fig:flowfields}(d)--(f) show the corresponding plots of the vertical velocity field $w$. Panels (a) and (d) of Figure~\ref{fig:flowfields} indicate that, under strong stratification ($\Fr = 0.004$), the flow organizes into layers. The $\theta(\mbfx)$ field exhibits bright yellow and dark spots associated with localized regions of enhanced and reduced density fluctuations, respectively. The layering depends on the stratification strength; increasing $\Fr$ leads to a reduction in layering and an increase in the characteristic vertical length scale. However, these structures remain stable because buoyancy forces dominate the nonlinear inertial terms and viscosity is sufficient to suppress secondary instabilities. With further increase in $\Fr$, the flow lies in a weakly turbulent, wave-dominated regime, characterized by a proliferation of localized turbulent structures, while viscosity limits the development of small-scale instabilities.

However, when $\Fr$ is increased to $0.038$, the flow develops overturning regions, as illustrated in Figure~\ref{fig:flowfields}(b). The stable density layering is disrupted, leading to the formation of roll structures characteristic of Kelvin–Helmholtz instabilities. A further weakening of stratification with increasing $\Fr$ leads to a significant reduction in layering; at $\Fr=0.051$, Figure~\ref{fig:flowfields}(c) shows that the layering disappears. This occurs when the nonlinear terms become significant compared with buoyancy forces and the flow transitions to a nearly isotropic turbulent regime, in which the scalar field $\theta(\mbfx)$ is passively advected.

\subsection{Transition to stratified turbulence}

\begin{figure*}
	\includegraphics[width=0.29\linewidth]{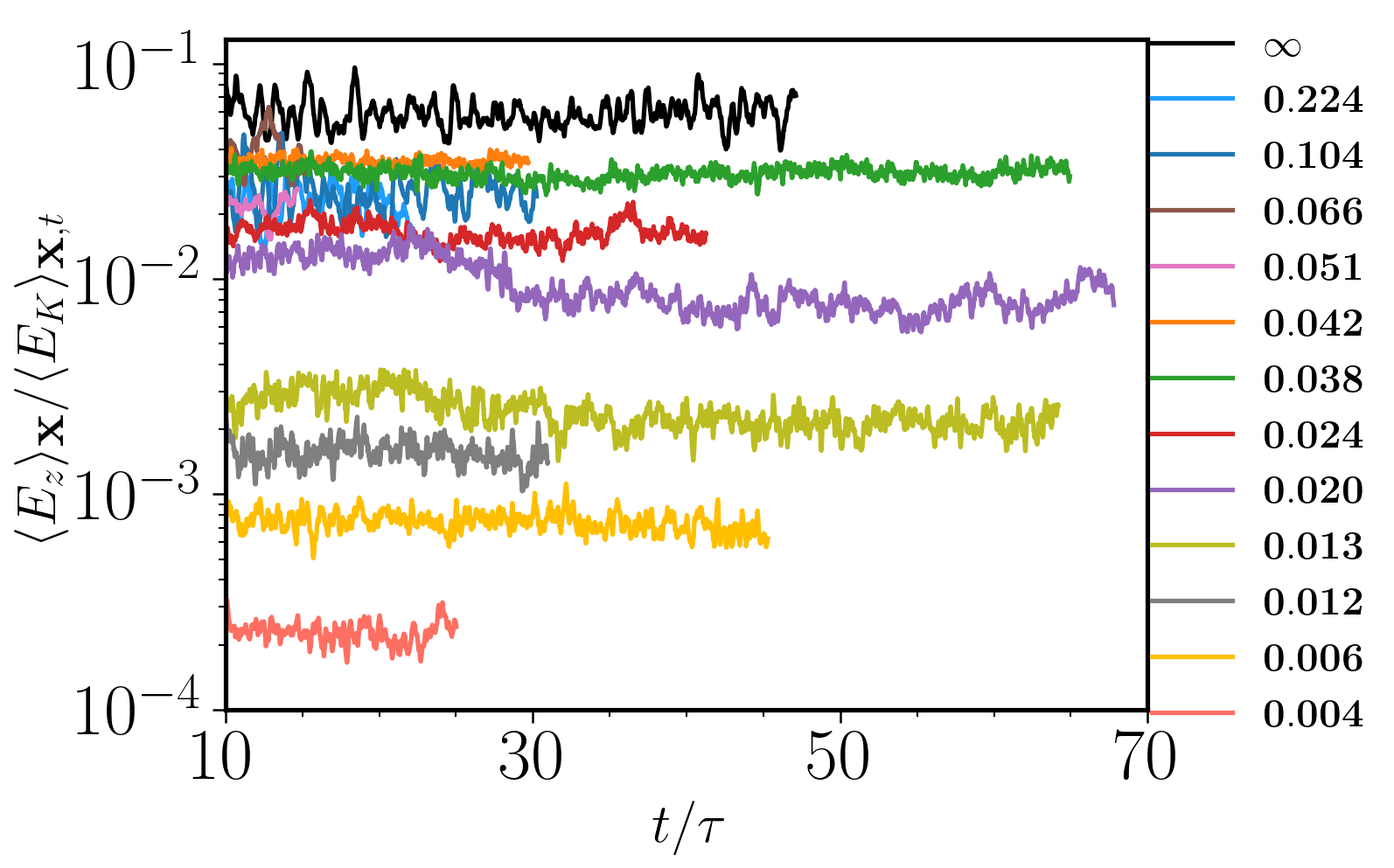}
	%\put(-75,130){{\large{(a)}}}
	\put(-65,32){{\large{(a)}}}
	\includegraphics[width=0.33\linewidth]{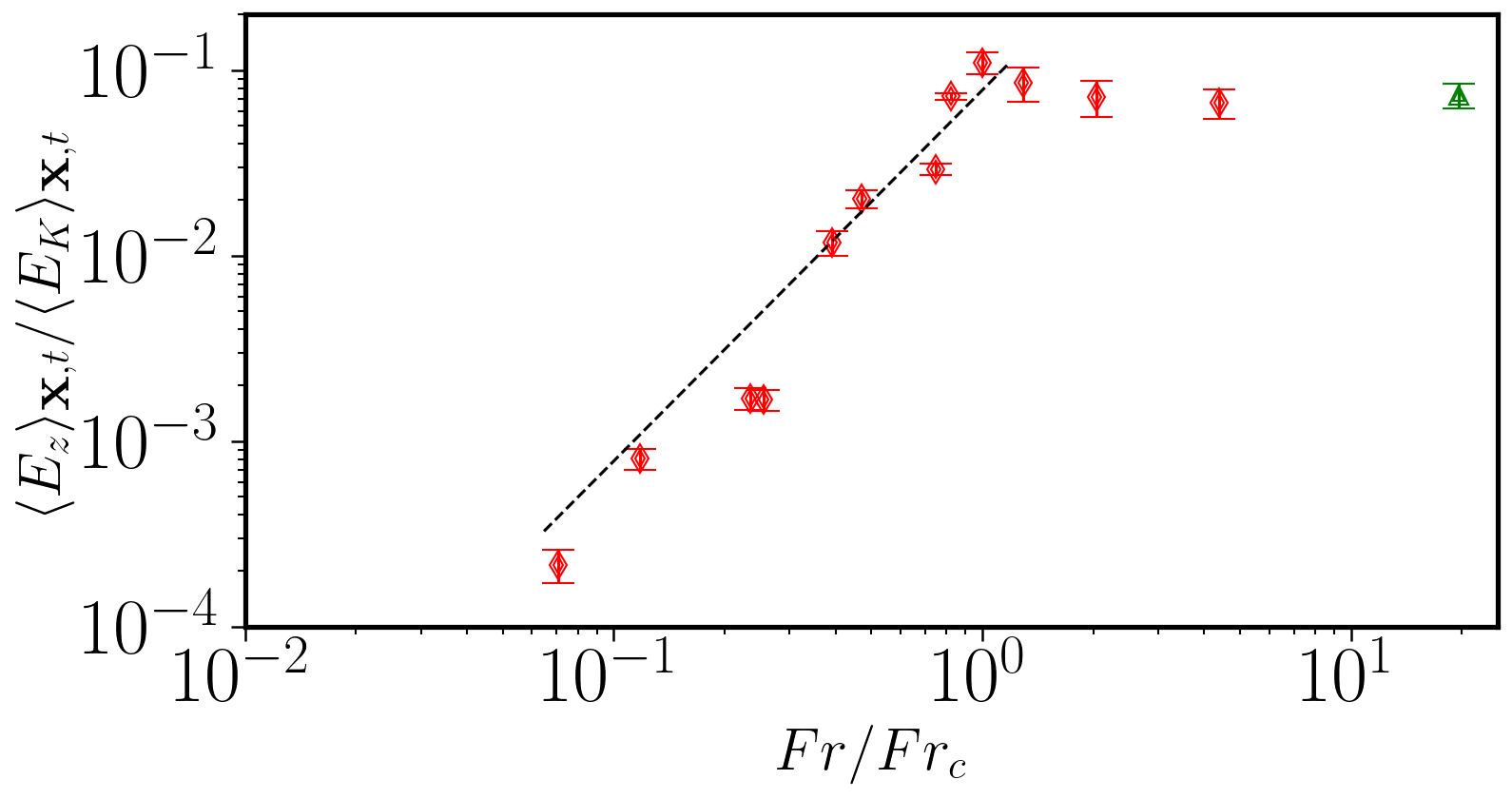}
	%\put(-140,33){\includegraphics[width=0.26\linewidth]{figures/LparayLperpLogLog}}
	\put(-70,32){{\large{(b)}}}
	\includegraphics[width=0.33\linewidth]{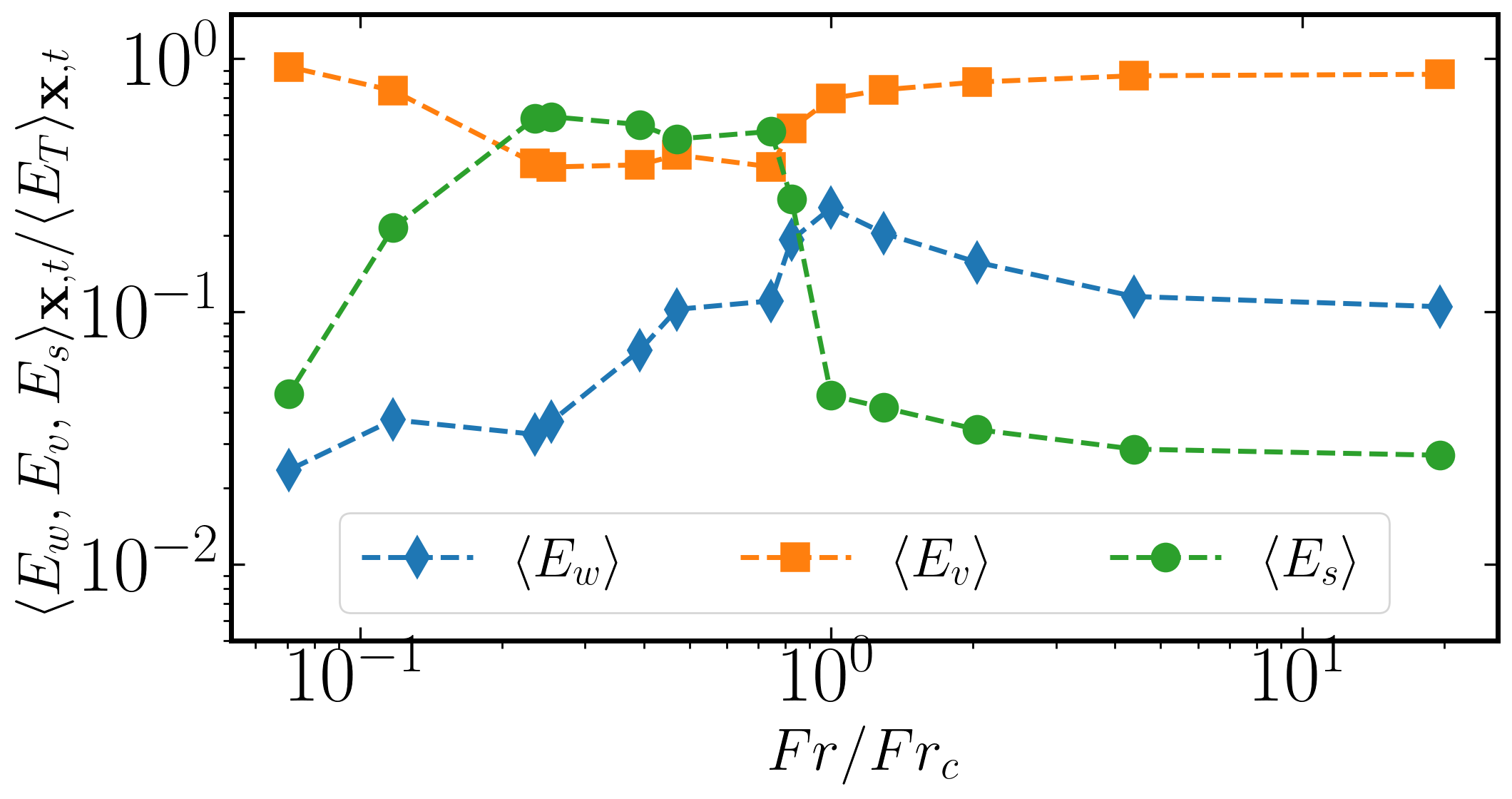}
	\put(-95,32){{\large{(c)}}}
	\caption{Vertical kinetic energy and energy partition among flow components.
		(a) Time evolution of the fraction of vertical kinetic energy, $\langle E_z \rangle_{\mathbf{x}}(t)/\langle E_K \rangle_{\mathbf{x},t}$, in the statistically steady state for different Froude numbers $\Fr$. The black line corresponds to the unstratified case ($\Fr=\infty$). Time is normalized by the large-eddy turnover time $\tau = L/U$.
		(b) Time-averaged fraction of vertical kinetic energy, $\langle E_z \rangle_{\mathbf{x},t}/\langle E_K \rangle_{\mathbf{x},t}$, versus the reduced Froude number $\Fr/\Fr_c$. Error bars indicate uncertainties equal to one standard deviation. The green triangle denotes the unstratified case ($\Fr=\infty$). The critical Froude number $\Fr_c=0.051$ marks the transition to a nearly isotropic turbulent regime.
		(c) Energy contributions of different modes versus $\Fr/\Fr_c$: wave-like modes ($E_w$, blue dashed line with diamonds), vortical modes ($E_v$, orange dashed line with squares), and vertically sheared horizontal modes ($E_s$, green dashed line with squares). The total energy $E_T$ is the sum of kinetic and potential energies.
	}
	\label{fig:ez}
\end{figure*}

Figures~\ref{fig:flowfields}(d)--(f) show that the vertical velocity component becomes increasingly suppressed with decreasing $\Fr$. To examine this further, Figure~\ref{fig:ez}(a) shows the time evolution of the fraction of vertical kinetic energy, in the statistically steady state, $\langle E_z \rangle_{\textbf{x}}(t)/\langle E_K \rangle_{\textbf{x},t}$ for different values of $\Fr$, ($\langle E_K \rangle_{\textbf{x},t}$ is the space and time-averaged total kinetic energy). We note that DNS runs for $0.013 \leq \Fr \leq 0.104$ exhibit relatively long transients before reaching a steady state (data not shown). For $\Fr \geq 0.051$, the time series of $E_z(t)$ for different runs lie close to one another and are bounded above by the $\Fr=\infty$ limit; in this limit, the energy associated with the vertical velocity component scales as $\langle E_z \rangle_{\textbf{x},t}^\infty \simeq \gamma^2 \langle E_K \rangle_{\textbf{x},t}$.

Figure~\ref{fig:ez}(b) shows the space and time-averaged vertical kinetic energy fraction, $\langle E_z \rangle_{\textbf{x},t}/\langle E_K \rangle_{\textbf{x},t}$, as a function of $\Fr/\Fr_c$, where $\Fr_c = 0.051$. For $\Fr < \Fr_c$, the vertical kinetic energy decreases monotonically with increasing stratification strength and scales as $\langle E_z \rangle_{\textbf{x},t}/\langle E_K \rangle_{\textbf{x},t} \sim \Fr^{\delta}$, with an exponent $\delta \approx 2$, consistent with dimensional-analysis estimates for $\Fr \ll 1$. However, for $\Fr \geq \Fr_c$, the fraction $\langle E_z \rangle_{\textbf{x},t}/\langle E_K \rangle_{\textbf{x},t}$ saturates at approximately $\gamma^2$, with $\Rey$ nearly constant and the domain aspect ratio fixed at $\gamma=1/4$. Thus, $\Fr_c$ clearly separates two distinct flow regimes, marking a transition in the vertical velocity component.

Next, we characterize these flow regimes by estimating the energies associated with the vortical and wave-like modes, obtained using a wave–eddy decomposition of the velocity field. Figure~\ref{fig:ez}(c) shows the fractions of the total kinetic energy carried by the vortical ($E_v$), wave-like ($E_w$), and VSHF ($E_s$) modes as functions of $\Fr/\Fr_c$. We find that $E_v$ dominates the energy budget and is nearly equal to the total energy $E_T (= E_K + E_P)$ for $\Fr > \Fr_c$. For $\Fr < \Fr_c$, $E_v$ drops rapidly to $E_v \approx 0.4 E_T$, remains nearly constant until $\Fr = 0.012$, and then increases gradually, approaching $E_T$ at $\Fr = 0.004$, the strongest stratification considered. The energy of the wave-like component, $E_w$, is more than an order of magnitude smaller than that of the vortical component for $\Fr \ll \Fr_c$. As $\Fr$ increases, $E_w$ grows approximately linearly, reaching a peak value of $E_w \approx 0.25 E_T$ at $\Fr \approx \Fr_c$, after which it gradually decreases to $E_w \approx 0.1 E_T$. Interestingly, the energy associated with the VSHF modes increases from a negligible value at $\Fr=0.004$ to $E_s \approx 0.6 E_T$ at $\Fr=0.012$, and remains approximately constant up to $\Fr \approx 0.04$. It then drops sharply, returning to a negligible value for $\Fr \gtrsim \Fr_c$. Thus, there exists a range of $\Fr < \Fr_c$ over which the VSHF modes dominate the energy budget, accompanied by a reduction in the energy of the vortical modes.

 \begin{figure*}
 	\includegraphics[width=0.25\linewidth]{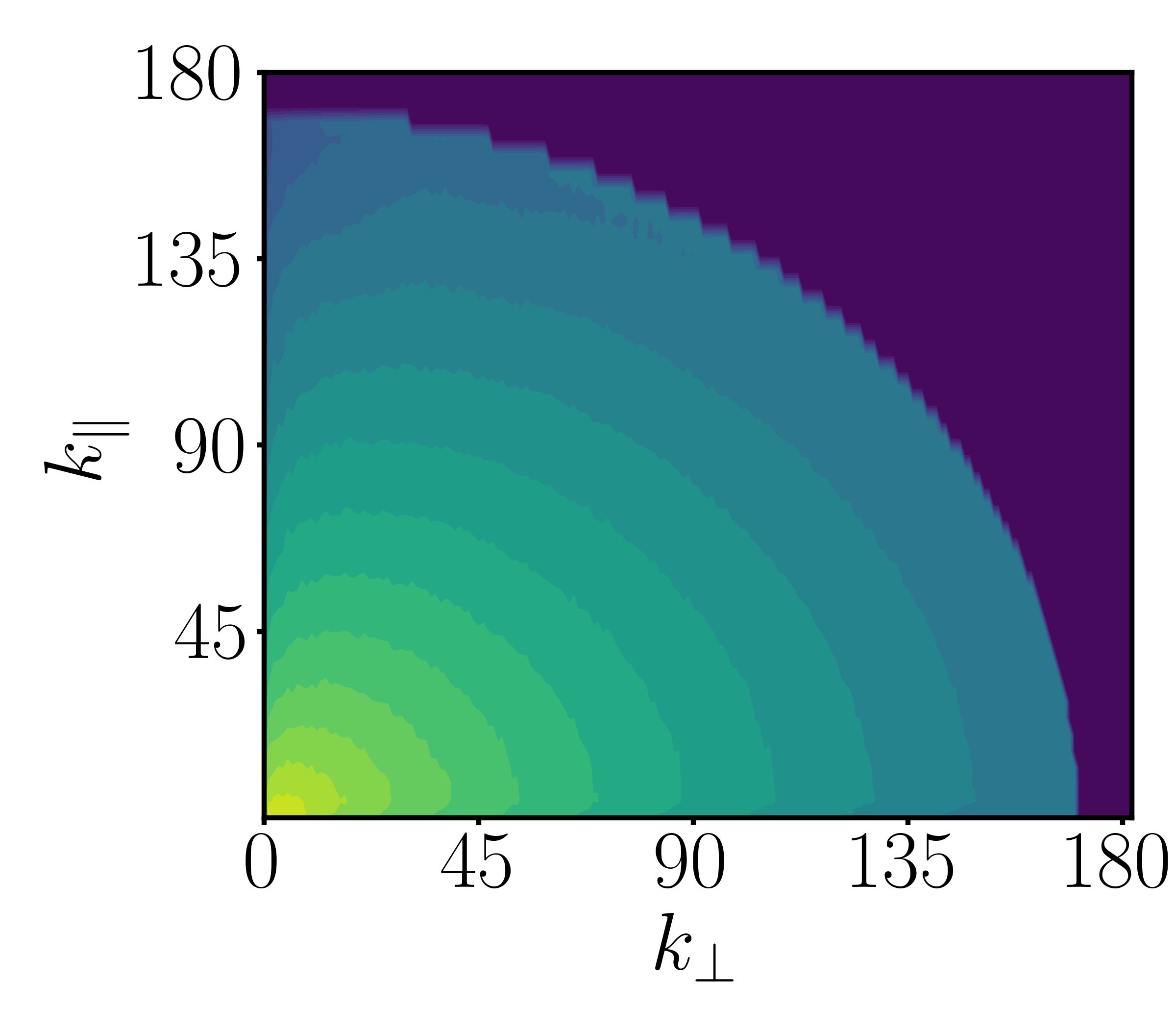}
 	\put(-60,25){{\large{(a)}}}
 	\includegraphics[width=0.24\linewidth]{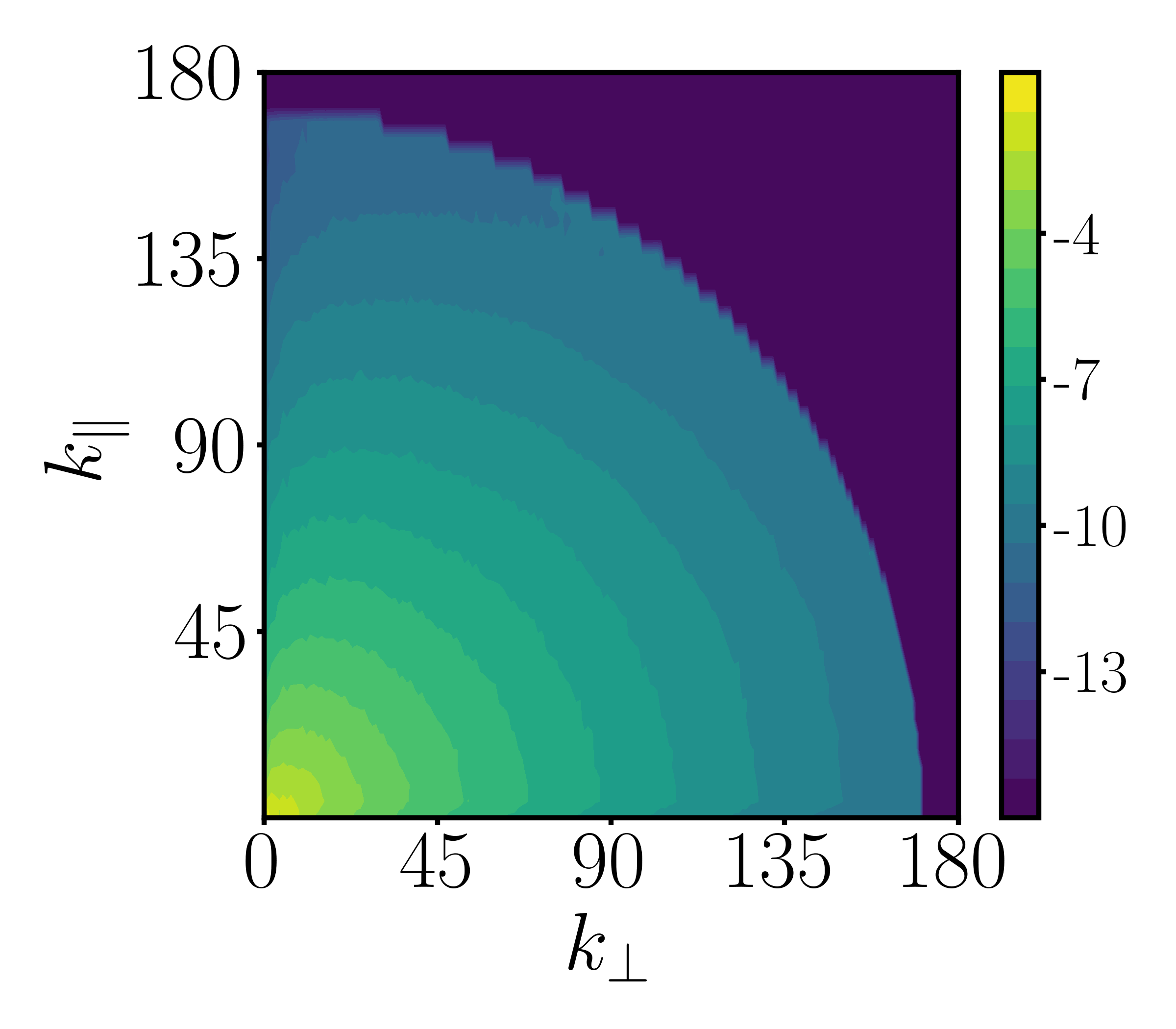}
 	\put(-70,25){{\large{(b)}}} 
 	\includegraphics[width=0.24\linewidth]{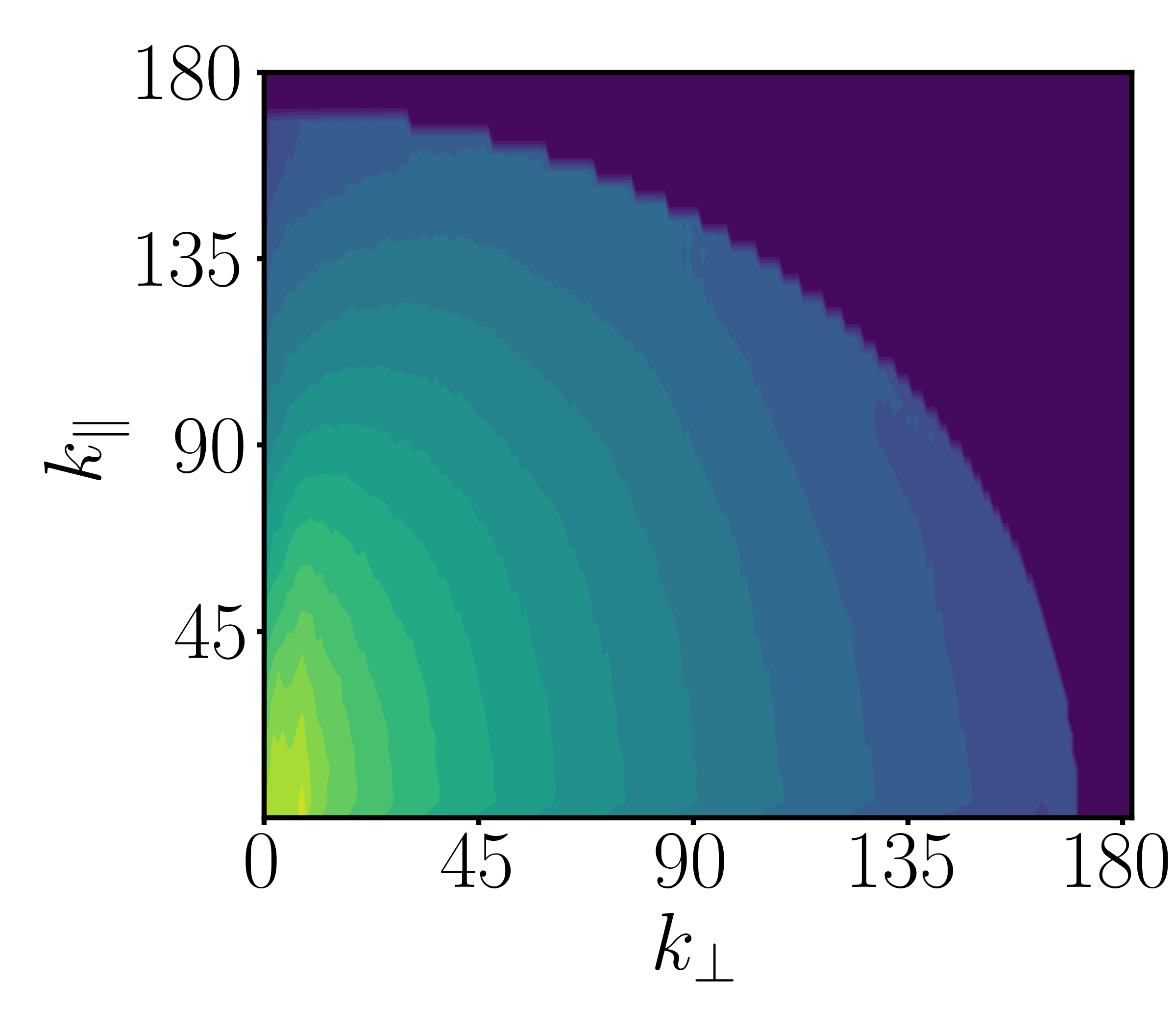}
 	\put(-70,25){{\large{(c)}}}
 	\includegraphics[width=0.25\linewidth]{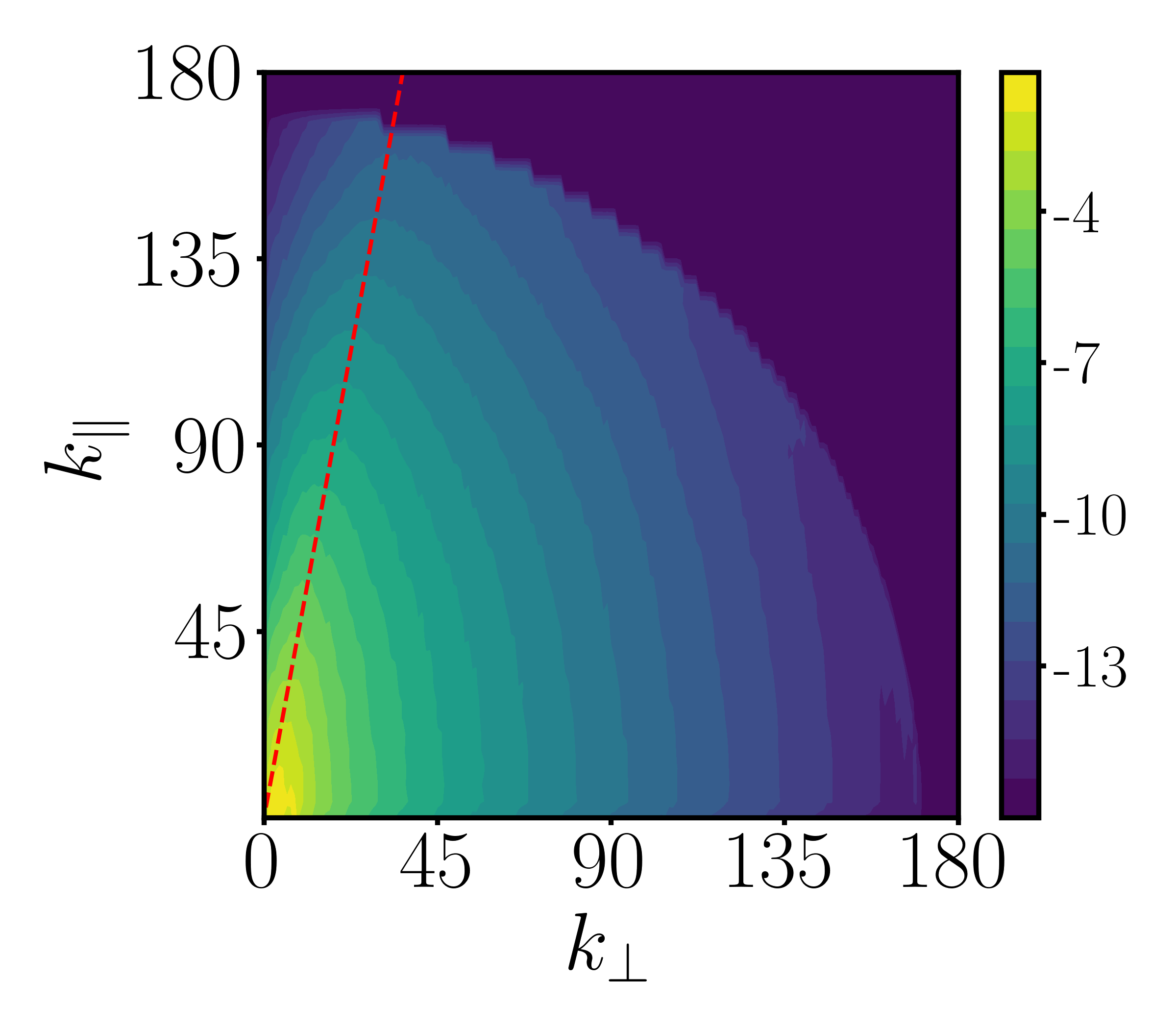}
 	\put(-80,25){{\large{(d)}}}
 	\caption{Axisymmetric kinetic energy spectra.
 		Pseudocolor plots of the two-dimensional axisymmetric kinetic energy spectra $E(k_{\perp},k_{\parallel})$ for different Froude numbers: (a) $\Fr=\infty$ (unstratified case), (b) $\Fr=0.104$ (very weak stratification), (c) $\Fr=0.013$ (moderate stratification), and (d) $\Fr=0.004$ (strong stratification). Here $k_{\perp}=\sqrt{k_x^2+k_y^2}$ and $k_{\parallel}=k_z$, where $k_i$ denotes the wave number along the $i$th Cartesian direction. The red dashed line marks the scaling $k_{\parallel} \approx 5 k_{\perp}$, revealing a pronounced concentration of energy along this direction in spectral space.
 		}
 	\label{fig:axisymspec}
 \end{figure*}

\subsection{Energy spectra and fluxes}

In Figure~\ref{fig:axisymspec}, we present the steady-state 2D axisymmetric kinetic energy spectra for different values of $\Fr$. Figure~\ref{fig:axisymspec}(a) shows that, in the absence of stratification, the energy-spectrum contours are nearly isotropic, with only small deviations near $k_{\perp}=0$. For $k_{\parallel} \sim 0$, distinct peaks appear near $k_{\perp}=0$ and at the forcing wave number $k_f$, where most of the kinetic energy is concentrated. A comparison of Figures~\ref{fig:axisymspec}(b)--(d), corresponding to $\Fr=0.104$, $0.013$, and $0.004$, respectively, indicates that the degree of anisotropy increases as $\Fr$ decreases. This anisotropy is more pronounced at smaller $k_{\perp}$, with the isocontours becoming increasingly elongated along $k_{\parallel}$ (i.e., in the vertical direction). Nevertheless, at low to moderate stratification strengths the energy remains concentrated in the wavenumber range $0 < k_{\perp} < k_f$ for small vertical wavenumbers $k_{\parallel}$. For the strongest stratification considered here ($\Fr=0.004$), the energy is clearly concentrated along the line $k_{\parallel} \approx 5 k_{\perp}$ (indicated by the red dashed line in Figure~\ref{fig:axisymspec}(d)), with increasing depletion near $k_{\perp} \to 0$ as $k_{\parallel}$ increases.

\begin{figure*}
	\includegraphics[width=0.31\linewidth]{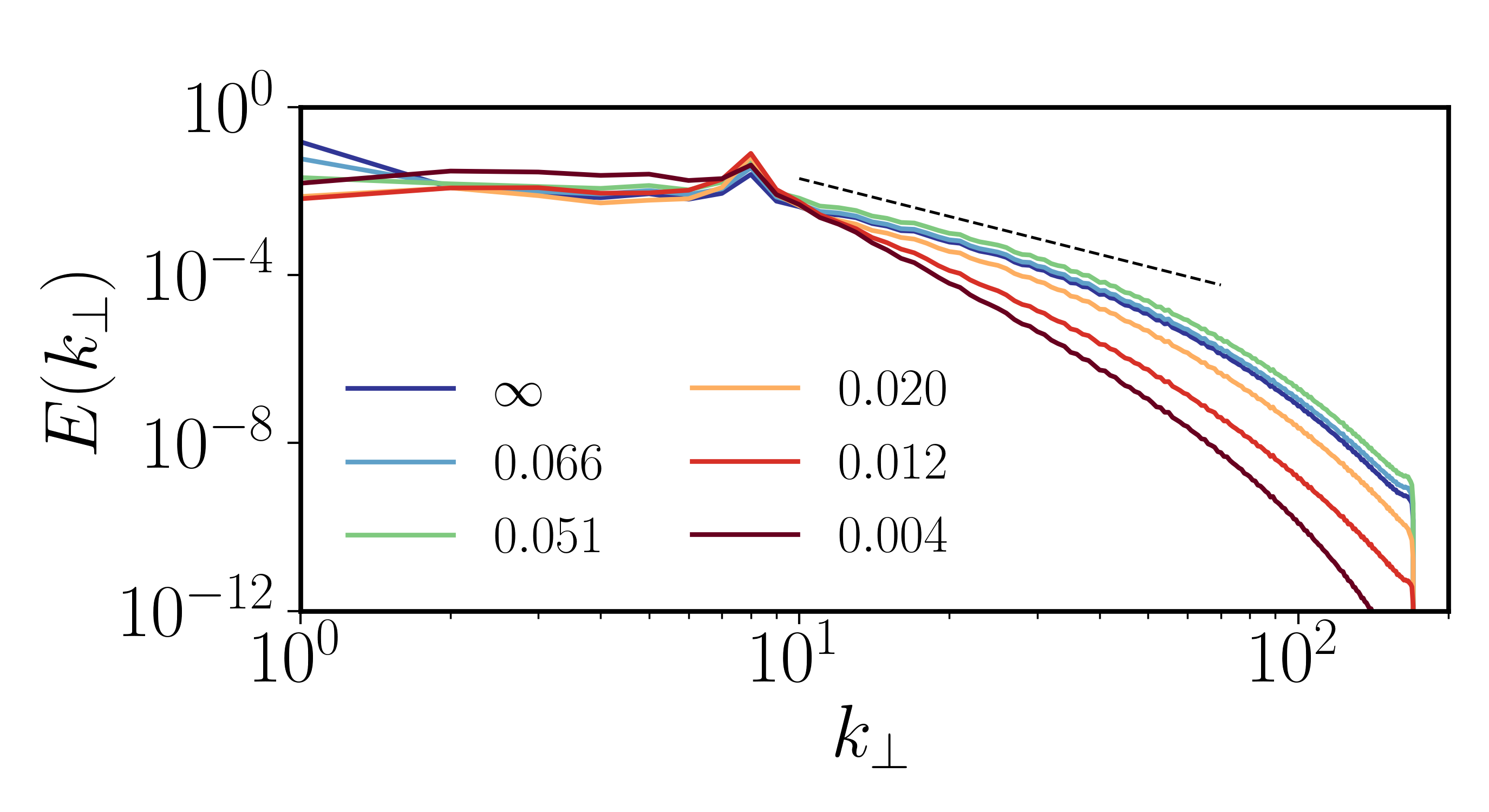}
	\put(-93,54){{\large{(a)}}}
	%\put(-200,25){\includegraphics[width=0.45\linewidth]{figures/labels}}
	\includegraphics[width=0.31\linewidth]{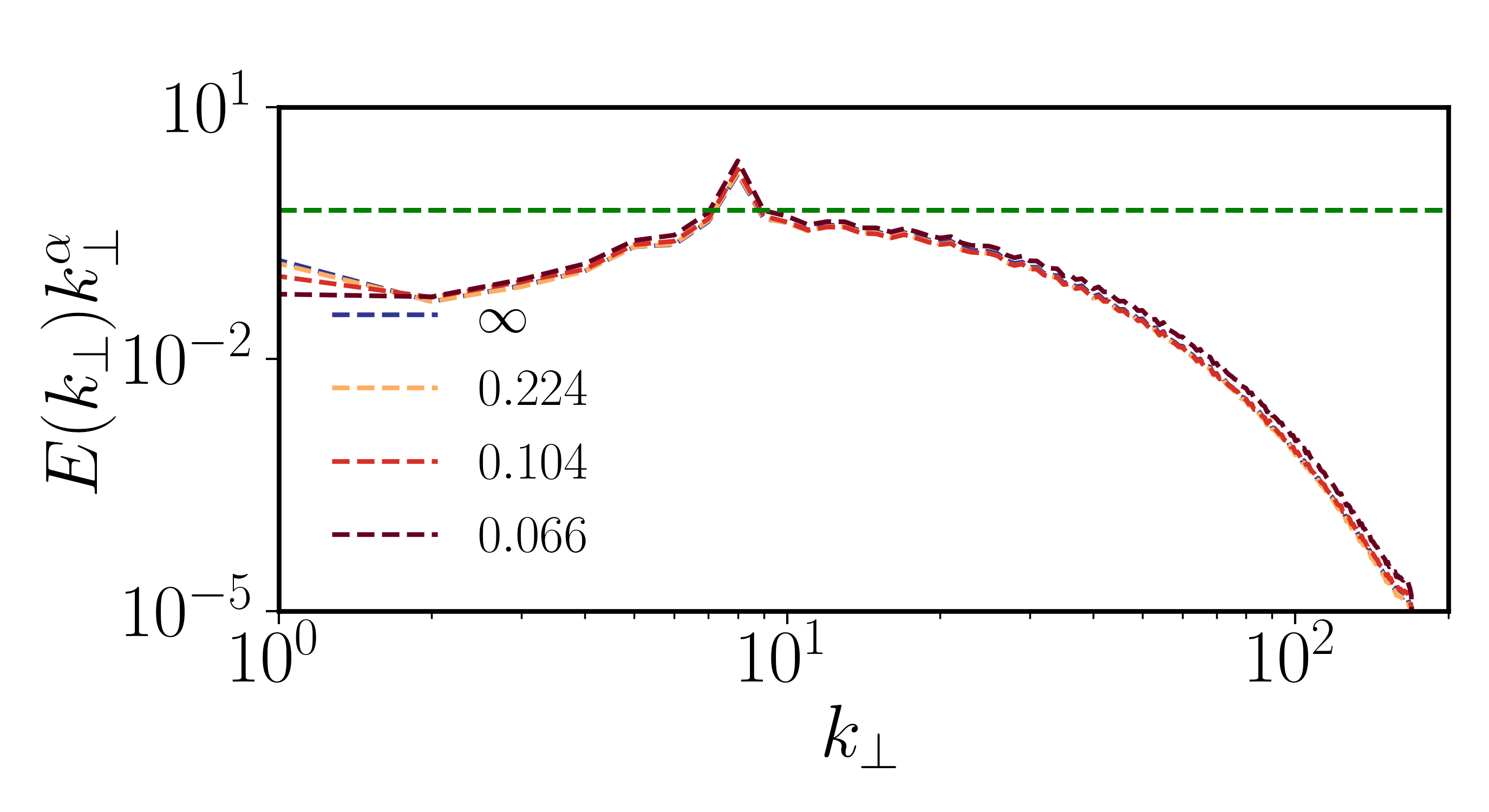}
	\put(-90,53){{\large{(b)}}}
	\includegraphics[width=0.31\linewidth]{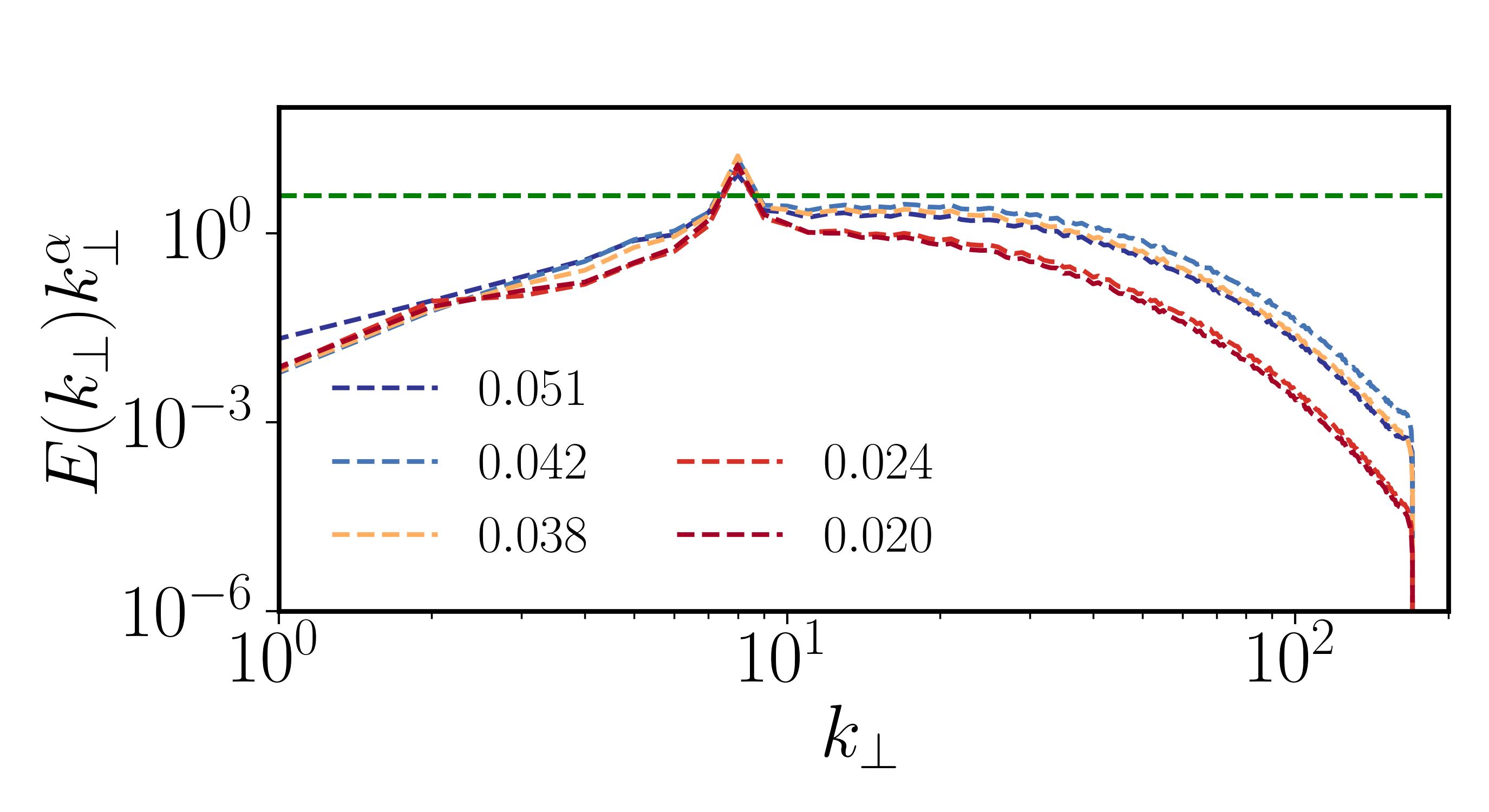}
	\put(-125,53){{\large{(c)}}}
	\\
	\includegraphics[width=0.31\linewidth]{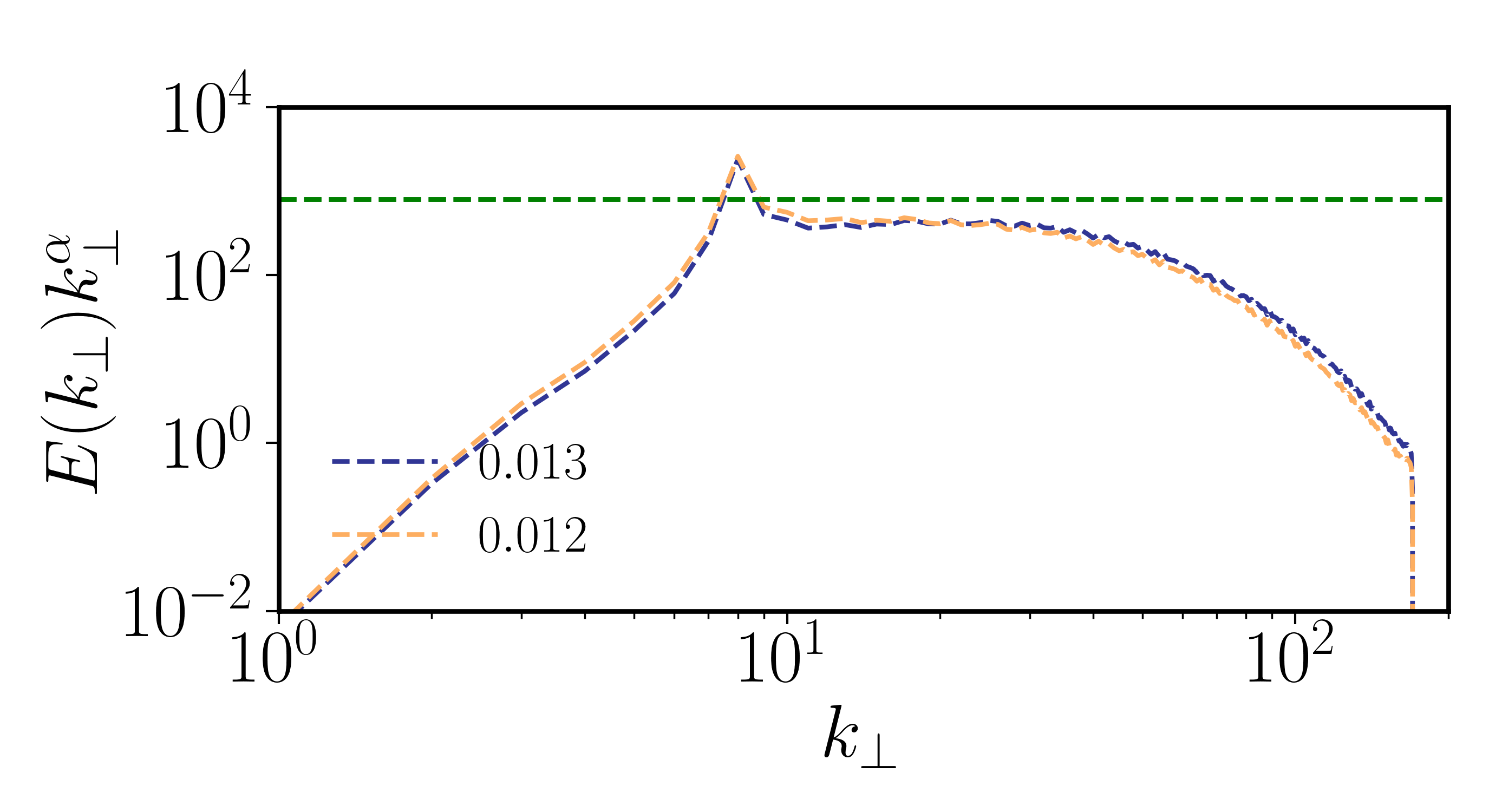}
	\put(-85,45){{\large{(d)}}}
	\includegraphics[width=0.31\linewidth]{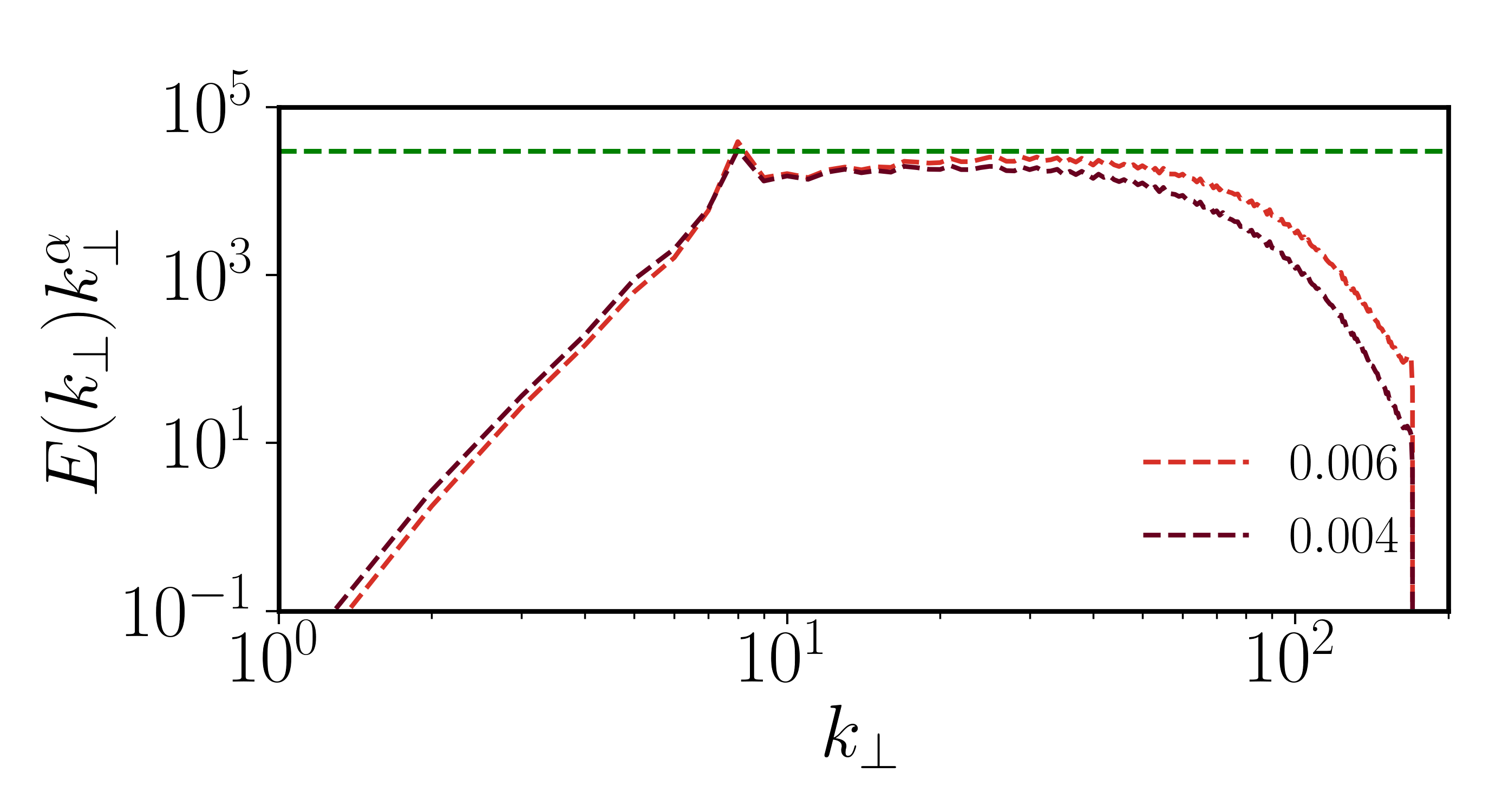}
	\put(-90,45){{\large{(e)}}}
	\includegraphics[width=0.31\linewidth]{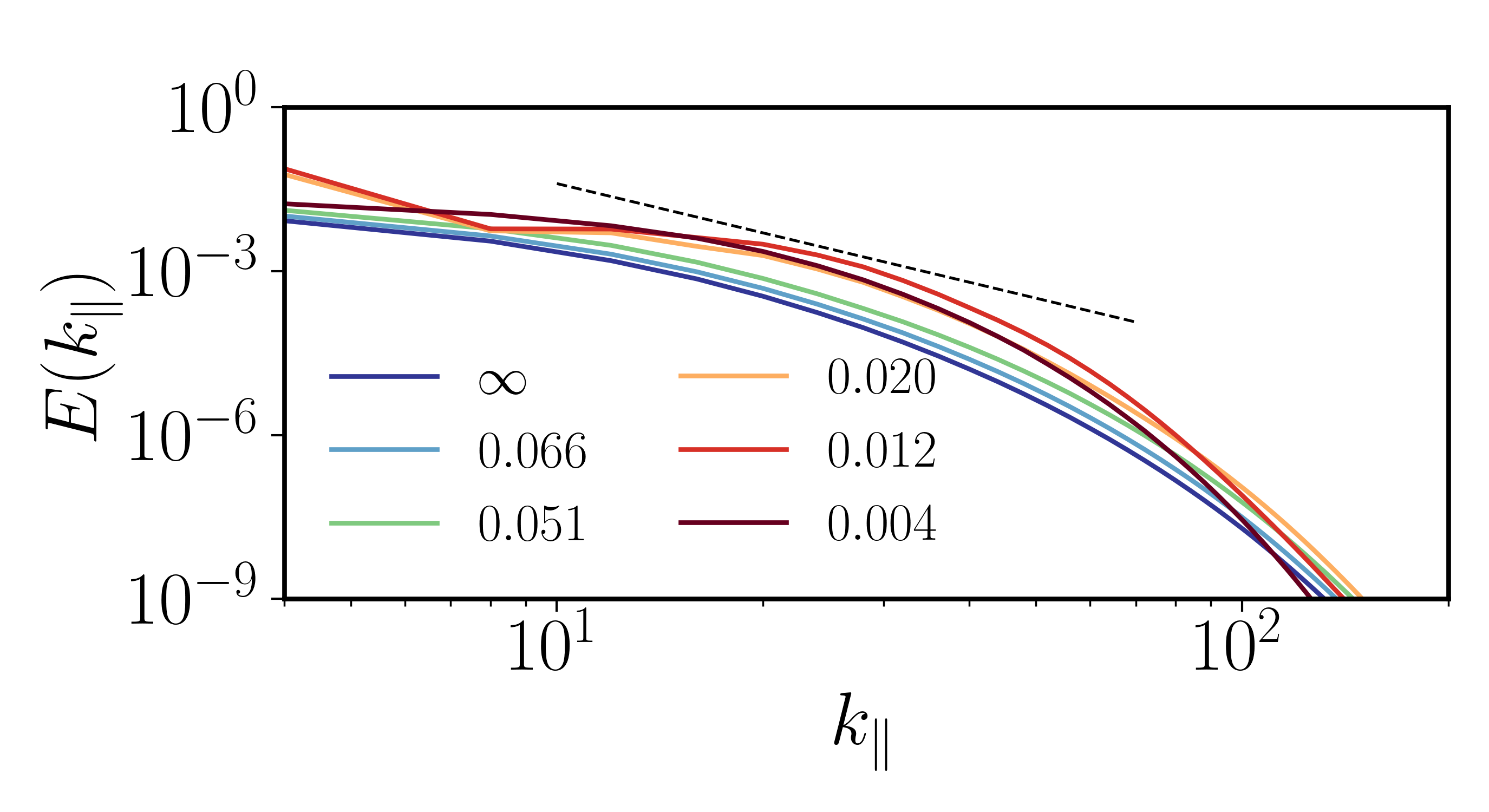}
	\put(-125,45){{\large{(f)}}}
	\caption{Energy spectra.
		(a) Reduced one-dimensional perpendicular kinetic energy spectra $E(k_{\perp})$ versus $k_{\perp}$ for different Froude numbers $\Fr$.
		(b–e) Compensated perpendicular spectra $k_{\perp}^{\alpha} E(k_{\perp})$ versus $k_{\perp}$ with $\alpha=2$ (b), $2.5$ (c), $5$ (d), and $6.5$ (e). The horizontal green dashed line indicates the corresponding $k_{\perp}^{-\alpha}$ scaling.
		(f) Reduced one-dimensional parallel kinetic energy spectra $E(k_{\parallel})$ versus $k_{\parallel}$ for different Froude numbers $\Fr$. The black dashed line indicates a $k_{\parallel}^{-3}$ scaling in (a,f).
		All spectra are time-averaged. $\Fr=\infty$ corresponds to the unstratified case.}
	\label{fig:spec}
\end{figure*}

We examine the time-averaged reduced 1D spectra, $E(k_{\perp})$ and $E(k_{\parallel})$, to understand how the scaling behaviour varies with stratification strength (Figure~\ref{fig:spec}). For $\Fr=\infty$ (deep blue curve) in Figure~\ref{fig:spec}(a), the spectrum $E(k_{\perp})$ is nearly flat at small wave numbers $k_{\perp} < k_f$, except for a peak at $k_{\perp}=1$. At larger wave numbers ($k_{\perp} > k_f$), a $k_{\perp}^{-2}$ scaling provides a reasonable fit, although the corresponding inertial range is rather limited, despite $\ell_f/\eta \approx 48$. We note that the Reynolds number in our DNS runs is relatively modest, $\Rey \sim 3000$. Introducing weak stratification ($\Fr = 0.224, 0.104, 0.066$) does not significantly alter the spectrum, aside from a slight reduction in the peak amplitude at $k_{\perp}=1$. A further increase in stratification ($\Fr=0.051$, $0.042$, $0.038$, $0.024$, and $0.02$) yields a flat spectrum for $k_{\perp}<k_f$, while for $k_{\perp}>k_f$ the perpendicular kinetic energy spectra are better described by a $k_{\perp}^{-2.5}$ scaling. This behaviour is illustrated by the compensated perpendicular kinetic energy spectra $k_{\perp}^{-\alpha}E(k_{\perp})$ shown in Figure~\ref{fig:spec}(b) and (c) for $\alpha=2$ and $2.5$, respectively. For $\Fr=0.013$ and $0.012$, the spectrum remains flat for $k_{\perp}<k_f$, whereas for $k_{\perp}>k_f$ a steeper $k_{\perp}^{-5}$ scaling provides an increasingly better fit, as shown in Figure~\ref{fig:spec}(d). Similarly, for $\Fr=0.006$ and $0.004$, Figure~\ref{fig:spec}(e) shows that $E(k_{\perp}) \sim k_{\perp}^{-6.5}$ over the range $k_f < k_{\perp} \lesssim 50$.

Figure~\ref{fig:spec}(f) shows that the parallel (vertical) kinetic energy spectrum, $E(k_{\parallel})$, does not exhibit a well-defined scaling region. Here, the buoyancy wave number $k_b$ plays an important role: the spectrum shows a tendency to flatten for $k_{\parallel} < k_b$, with $k_b$ increasing as $\Fr$ decreases (see Table~\ref{tab:runs}). For $k_{\parallel} > k_b$, $E(k_{\parallel})$ appears to decay more steeply than $k_{\parallel}^{-3}$, as indicated by the dashed black line. We recall that for $\Fr < 0.013$, the available wave number range between $k_b$ and the dissipation wave number $k_{\eta}$ is less than a decade, with $k_{\eta}/k_b \approx 1.7$ at $\Fr = 0.004$, while the Ozmidov wave number lies well within the dissipation range.

\begin{figure}
	\includegraphics[width=0.45\linewidth]{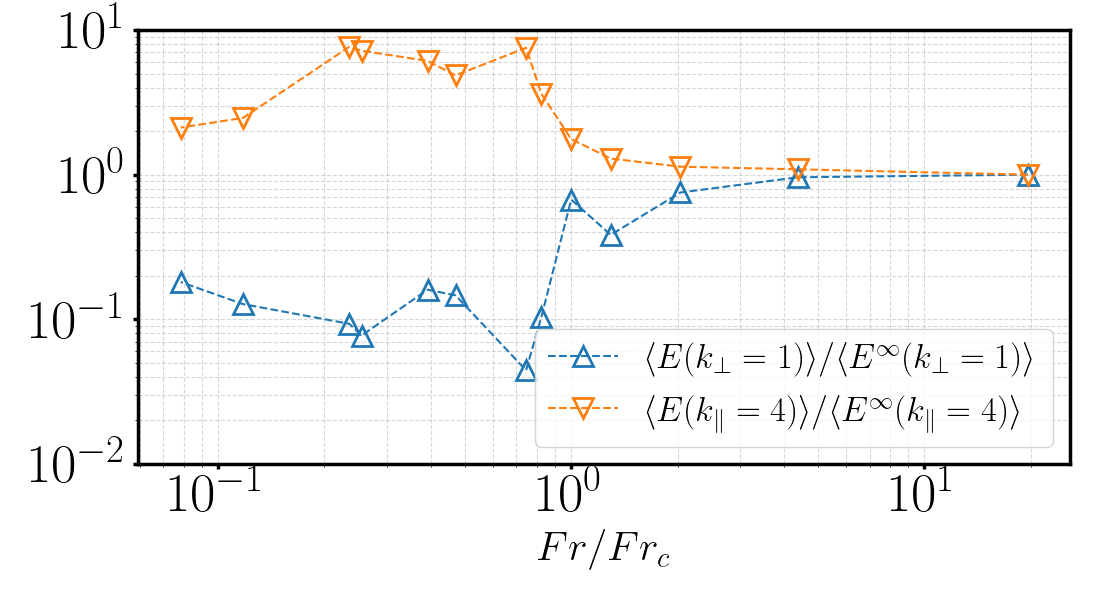}
	\put(-28,100){{\large{(a)}}}
	\includegraphics[width=0.45\linewidth]{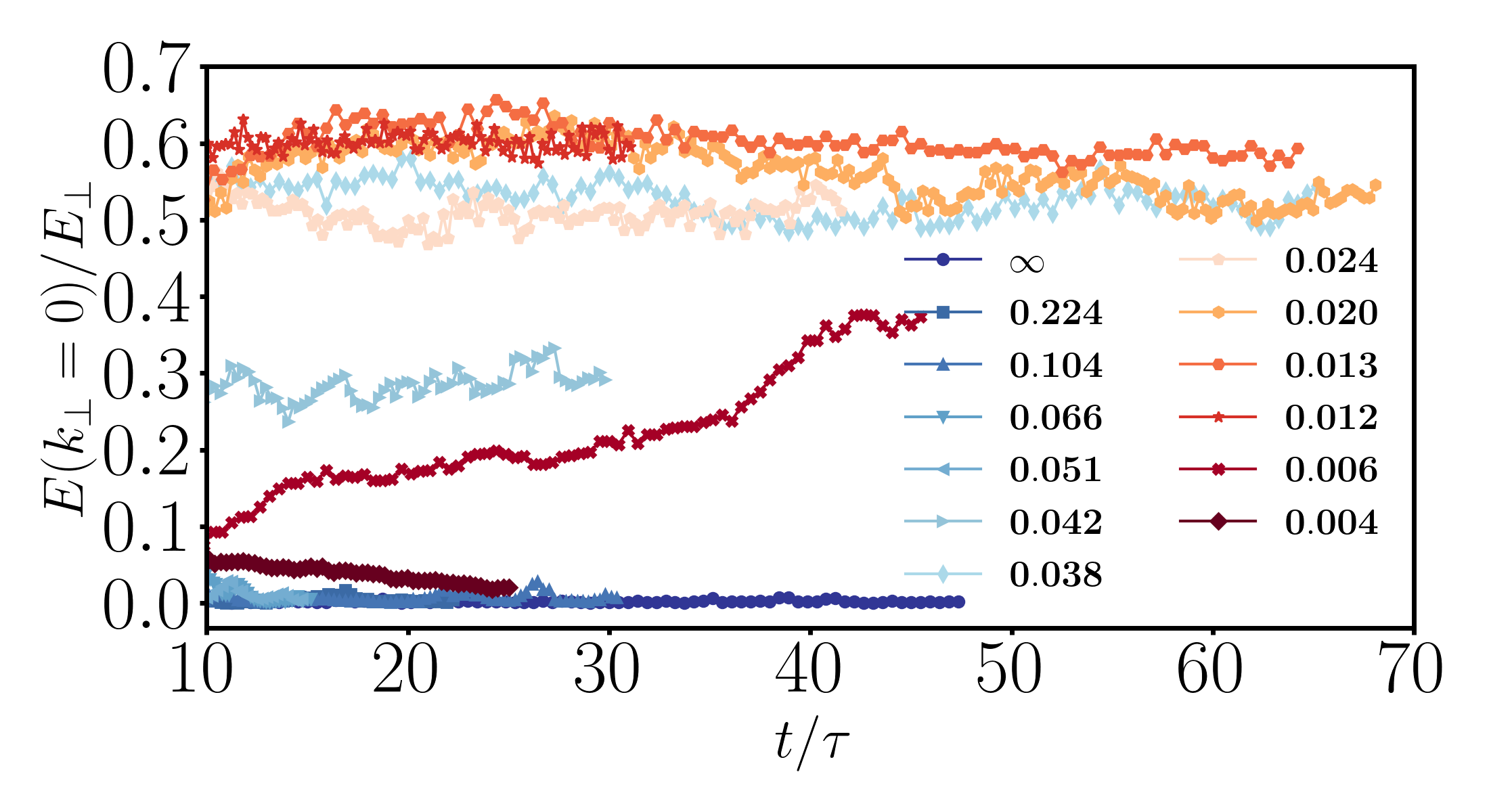}
	\put(-28,100){{\large{(b)}}}
	\caption{Kinetic energy at low wave numbers and VSHFs.
		(a) Time-averaged energy content $\langle E(k_{\perp}=1)\rangle_t/ \langle E^{\infty}(k_{\perp}=1)\rangle_t$ and $\langle E(k_{\parallel}=4)\rangle_t / \langle E^{\infty}(k_{\parallel}=4)\rangle_t$ versus the reduced Froude number $\Fr/\Fr_c$. Here, $k_{\perp}=1$ (= $k_{\perp}^{\min}$) and $k_{\parallel}=4$ (= $k_{\parallel}^{\min}$) are the smallest non-zero wave numbers in the horizontal and vertical directions, respectively. The quantities $E^{\infty}(k_{\perp}^{\min})$ and $E^{\infty}(k_{\parallel}^{\min})$ denote the corresponding values in the absence of stratification.
		(b) Energy in the vertically sheared horizontal flow (VSHF) modes, $E(k_{\perp}=0)/E_{\perp}$, versus $t/\tau$, where $E_{\perp}$ denotes the total kinetic energy in the perpendicular modes and $\tau$ is the large-eddy turnover time. The critical Froude number $\Fr_c=0.051$ marks the transition to a nearly isotropic turbulent regime. 
	}
	\label{fig:ekperpk1parak4_avgN}
\end{figure}

Both the perpendicular and parallel energy spectra in Figure~\ref{fig:spec} indicate that the energy content at the lowest wave numbers depends on $\Fr$. To illustrate this, Figure~\ref{fig:ekperpk1parak4_avgN}(a) shows the time-averaged values of $E(k_{\perp}=1)$ and $E(k_{\parallel}=4)$, normalized by their respective values in the absence of stratification ($\Fr=\infty$). Figure~\ref{fig:ekperpk1parak4_avgN}(a) shows that $\langle E(k_{\perp}=1)\rangle_t/ \langle E^{\infty}(k_{\perp}=1) \rangle_t$ and $\langle E(k_{\parallel}=4)\rangle_t/ \langle E^{\infty}(k_{\parallel}=4) \rangle_t$ approach unity as $\Fr \to \infty$, indicating that the perpendicular and parallel components attain their largest and smallest values, respectively, in the absence of stratification. The time-averaged $\langle E(k_{\perp}=1) \rangle_t$ decreases gradually with decreasing $\Fr$ until $\Fr_c$, where it drops by approximately an order of magnitude and approaches a plateau, before exhibiting a slight increase at stronger stratification (smaller $\Fr$). In contrast, the time-averaged parallel component $\langle E(k_{\parallel}=4) \rangle_t$ exhibits the opposite trend. We also note that $E(k_{\parallel}=4)$ generally requires a longer time to reach a statistically steady state than $E(k_{\perp}=1)$. For some intermediate values of $\Fr$, no clear saturation is observed over the full simulation duration (time series not shown).

Figure~\ref{fig:ekperpk1parak4_avgN}(b) shows the temporal evolution of the energy of VSHF modes, $E(k_{\perp}=0)$, for different stratification strengths. While these modes remain negligible at both weak stratification ($\Fr=0.224, 0.104, 0.066$, and $0.051$) and strong stratification ($\Fr=0.004$), they saturate at intermediate stratification strengths, accounting for approximately $50$--$60\%$ of the total kinetic energy in the perpendicular modes, $E_{\perp}$. We note that in Figure~\ref{fig:ekperpk1parak4_avgN}(b) the energy of the VSHF modes is normalized by the total kinetic energy of the perpendicular modes, whereas in Figure~\ref{fig:ez}(c) it is normalized by the total energy $E_T$.

\begin{figure*}
	\includegraphics[width=0.45\linewidth]{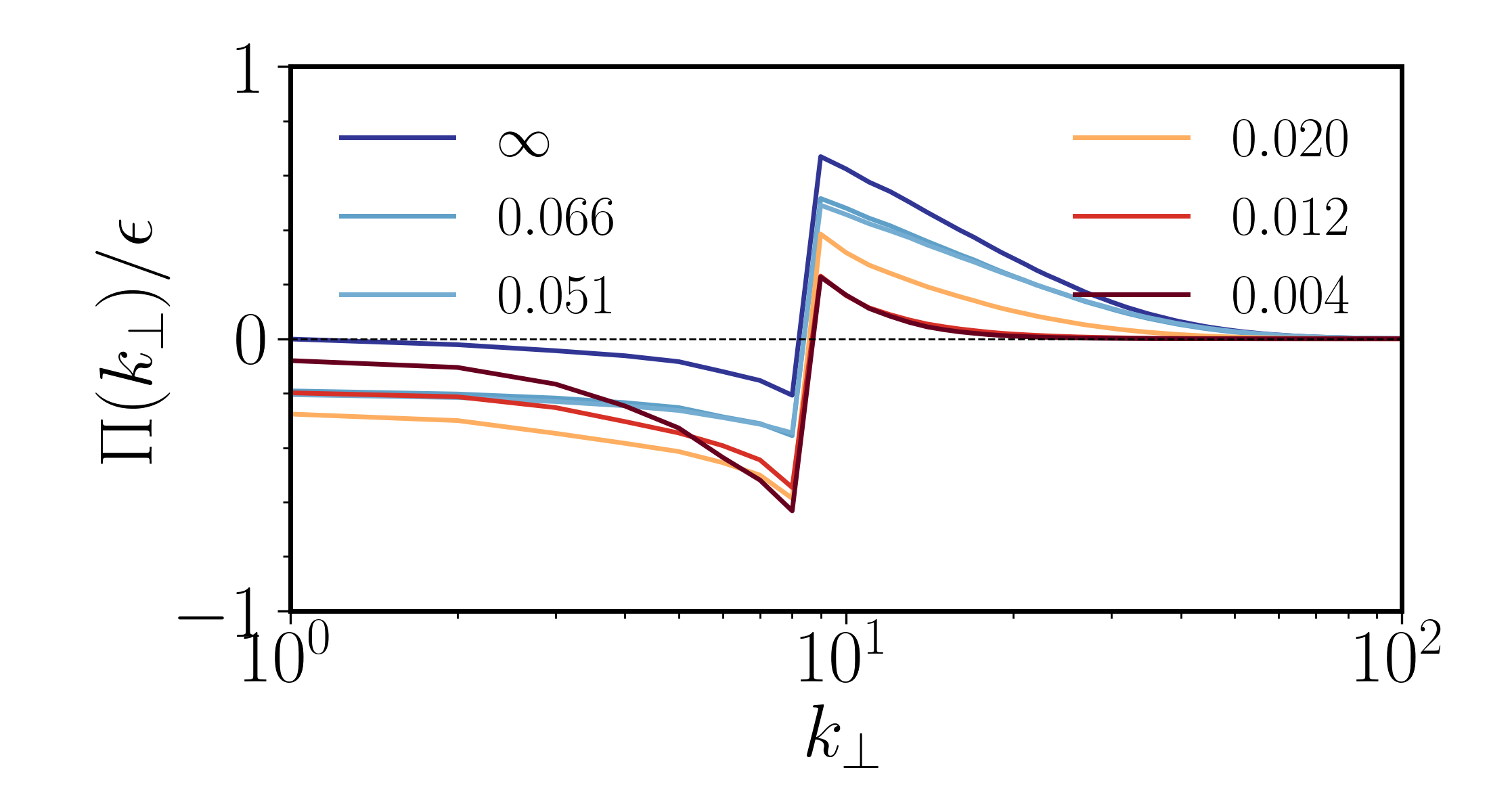}
	\put(-180,43){{\large{(a)}}}
	\includegraphics[width=0.45\linewidth]{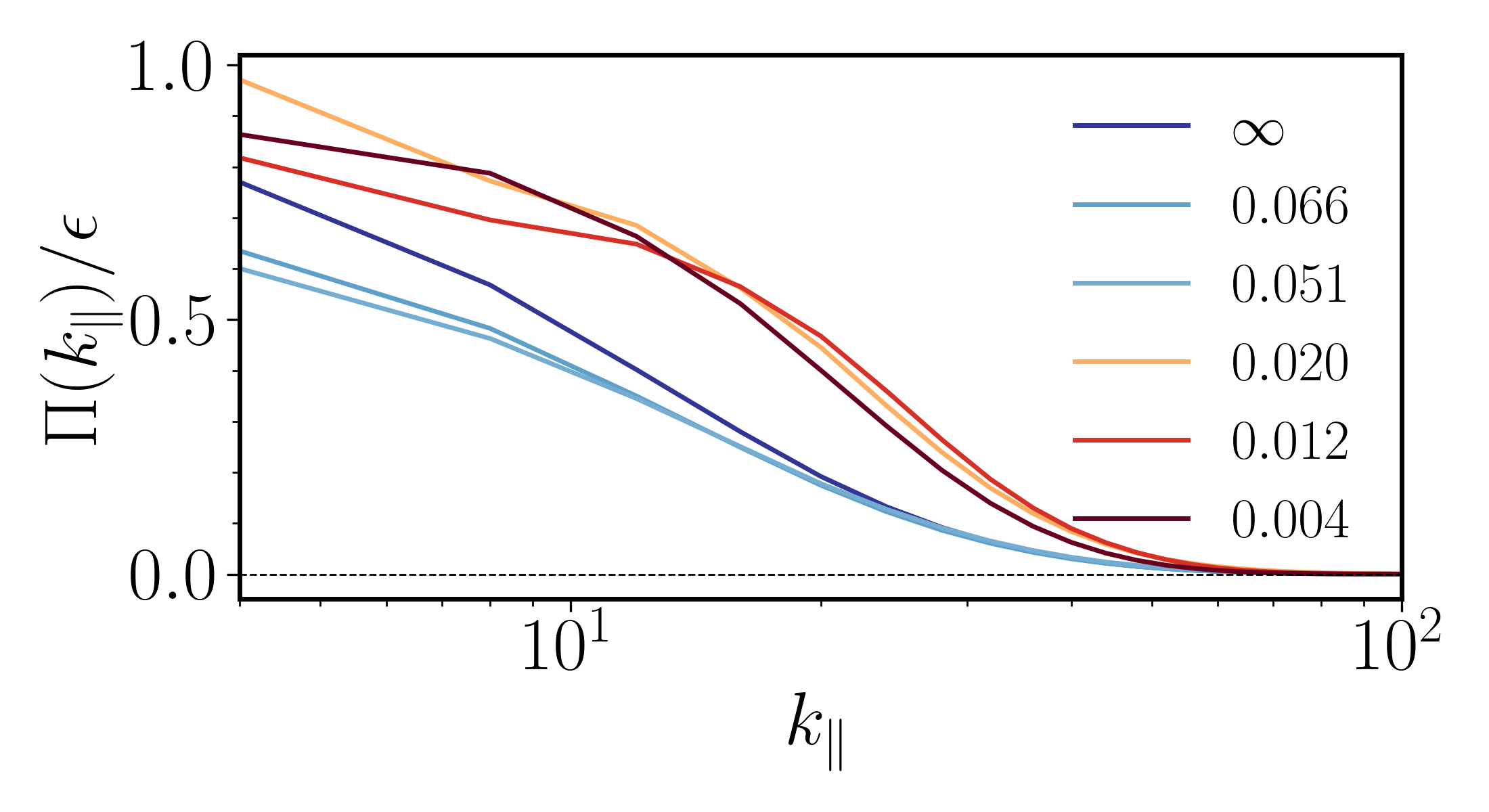}
	\put(-178,43){{\large{(b)}}}
	\caption{Energy fluxes.
		(a) Time-averaged perpendicular kinetic energy flux, $\Pi(k_{\perp})$, versus $k_{\perp}$.
		(b) Time-averaged parallel kinetic energy flux, $\Pi(k_{\parallel})$, versus $k_{\parallel}$.
		Fluxes are normalized by the kinetic energy injection rate $\epsilon$. The legend entries indicate the corresponding Froude numbers.
	}
	\label{fig:flux}
\end{figure*}

Figures~\ref{fig:flux}(a) and (b) show the perpendicular and parallel kinetic energy fluxes, $\Pi(k_{\perp})$ and $\Pi(k_{\parallel})$, respectively. We observe that $\Pi(k_{\perp})$ is negative for $k_{\perp} < k_f$, whereas $\Pi(k_{\parallel})$ remains positive over the entire accessible range of $k_{\parallel}$. This suggests that the nonlinear triadic interactions preferentially transfer energy toward larger horizontal length scales, while maintaining a forward transfer of energy in the vertical direction. This behavior is consistent with the characteristics of stratified flows, where energy preferentially accumulates at large horizontal length scales. The formation of a quasi-2D flow, comprising coherent VSHFs, acts as a sink that drains horizontal energy from the intermediate $k_{\perp}$ modes. The apparent negative perpendicular flux for $k_{\perp} < k_f$ does not represent a true inverse energy cascade, but instead reflects the anisotropic distribution of energy. The positive flux in the parallel direction ensures that the net energy transfer remains positive over this range of wave numbers. Moreover, for $k_{\perp} < k_f$, $\Pi(k_{\perp})$ varies non-monotonically with $\Fr$. At low wave numbers ($k_{\perp} \sim 1$), $\Pi(k_{\perp})$ plateaus at a finite but $\Fr$-dependent value, as shown in Figure~\ref{fig:flux}(a), whereas for $k_{\perp} > k_f$, the strength of the forward cascade decreases monotonically with decreasing $\Fr$.

\subsection{Vertical transport, intermittency, and mixing}

\begin{figure*}
	\includegraphics[width=0.48\linewidth]{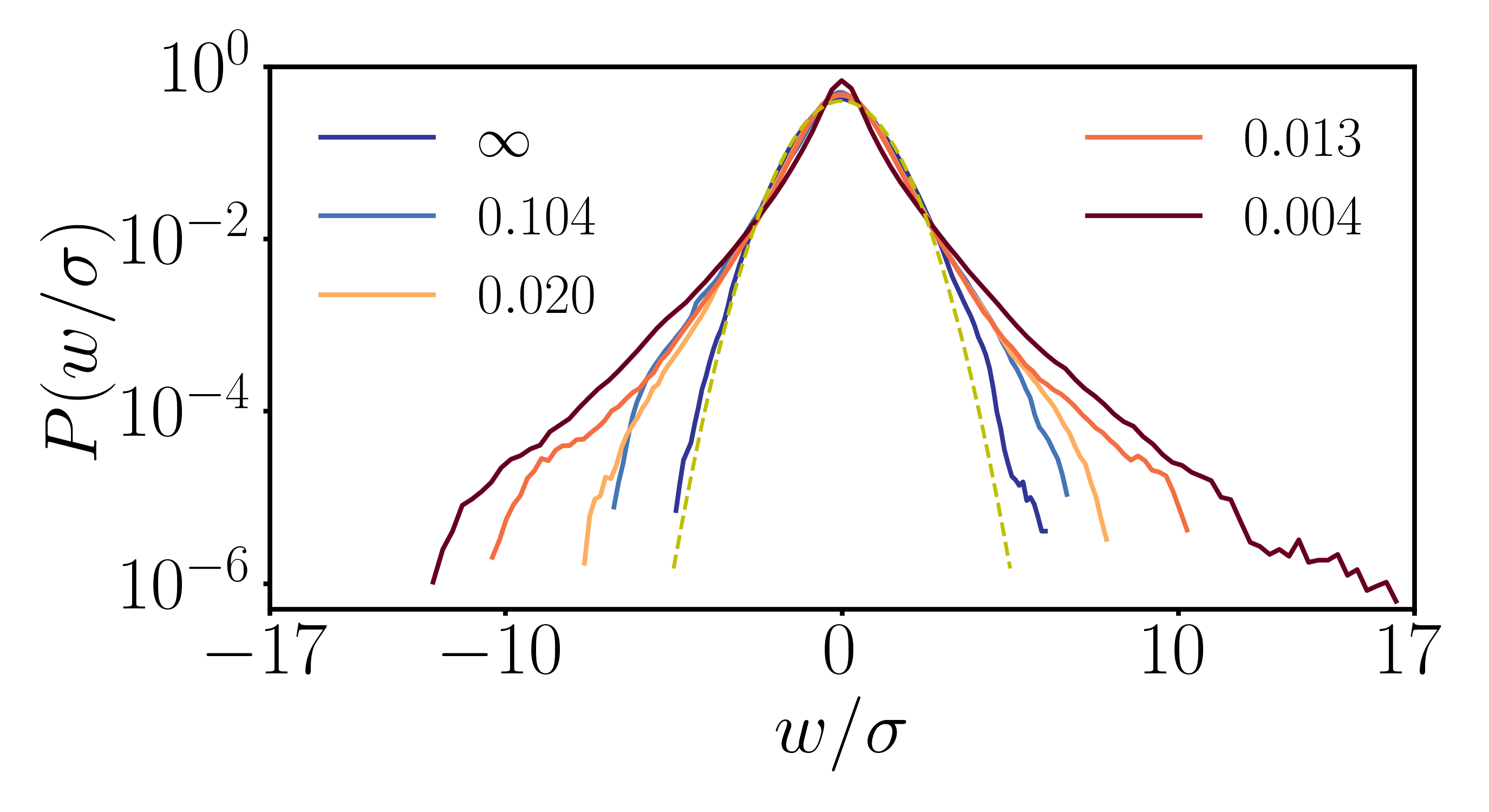}
	\put(-140,105){{\large{(a)}}}
	\includegraphics[width=0.47\linewidth]{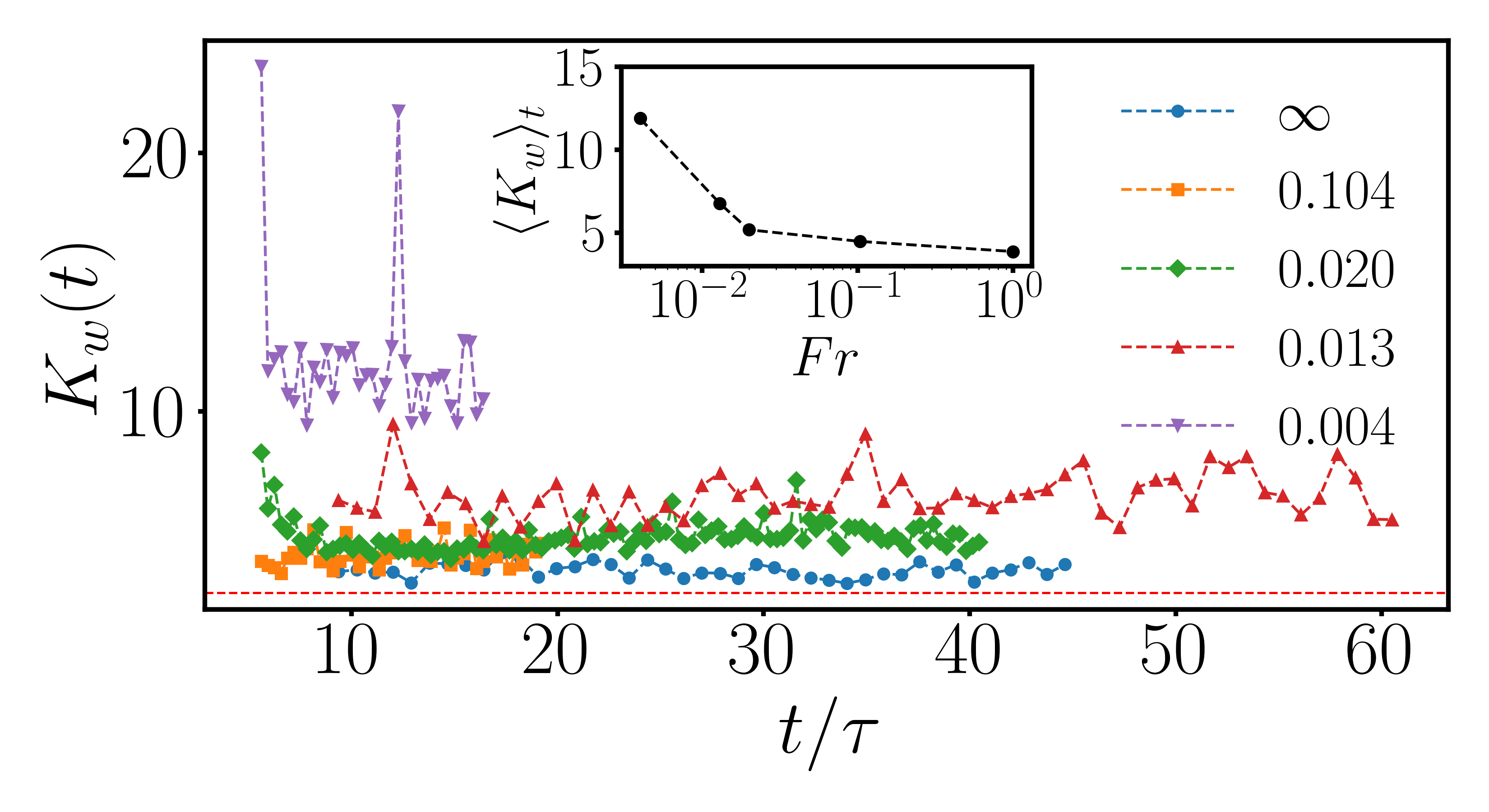}
	\put(-177,105){{\large{(b)}}}
	\caption{Intermittent vertical transport under stratification. (a) Probability distribution functions (PDFs) of the instantaneous vertical velocity field $w$, normalized by its standard deviation $\sigma$. The deep-blue solid line corresponds to the unstratified case, and the greenish-yellow dashed curve denotes a Gaussian distribution. The subscript $w$ on $\sigma$ is omitted here for notational simplicity. (b) Kurtosis of $w$ versus $t/\tau$, where $\tau$ is the large-eddy turnover time. The horizontal dashed line at $K_w=3$ indicates the Gaussian value for reference. Inset: time-averaged kurtosis $\langle K_w \rangle_t$ versus the Froude number $\Fr$, with $\Fr=\infty$ denoting the unstratified case.}
	\label{fig:wpdf}
\end{figure*}

To further characterize vertical transport, we examine the PDFs of the instantaneous vertical velocity field $w(\mbfx,t)$, shown in Figure~\ref{fig:wpdf}(a). At small and moderate stratification strengths, the PDFs are symmetric and approximately Gaussian (yellow dashed curve) near the centre, although their tails are relatively broad. At large stratification strengths, however, the PDFs become distinctly non-Gaussian and significantly broader; for example, the brown curve corresponding to $\Fr=0.004$ illustrates this pronounced deviation. To better quantify the tailedness of these distributions, we compute the kurtosis, defined as
\begin{equation}
	K_w(t) = \frac{\langle (w - \langle w \rangle_\mbfx)^4 \rangle_{\mbfx}}{\langle (w - \langle w \rangle_{\mbfx})^2\rangle_{\mbfx}^2}.
\end{equation}
Figure~\ref{fig:wpdf}(b) shows the temporal evolution of the kurtosis, $K_w(t)$, for $\Fr=\infty$, $0.104$, $0.024$, $0.013$, and $0.004$. The horizontal dashed line at $K_w=3$ indicates the Gaussian value for reference. Clearly, $K_w(t)$ deviates from this value, and both the degree of deviation and the magnitude of fluctuations increase as $\Fr$ decreases.

The inset of Figure~\ref{fig:wpdf}(b) shows the variation of the time-averaged kurtosis, $\langle K_w \rangle_t$, as a function of $\Fr$. For convenience, the $\Fr=\infty$ value is shown at $\Fr=1$, since all other $\Fr$ values considered are considerably smaller. We find that $\langle K_w \rangle_t$ decreases with increasing $\Fr$. Interestingly, the strongest stratification case ($\Fr=0.004$) exhibits the largest kurtosis, $\langle K_w \rangle_t \approx 12$, indicating highly intermittent vertical motions dominated by intense, localized bursts. Moreover, the time series of $K_w(t)$ for $\Fr=0.02$ and $0.004$ show that extreme events, with values reaching approximately $2\langle K_w \rangle_t$, can occur, although they are relatively infrequent.

\begin{figure}
	\includegraphics[width=0.4\linewidth]{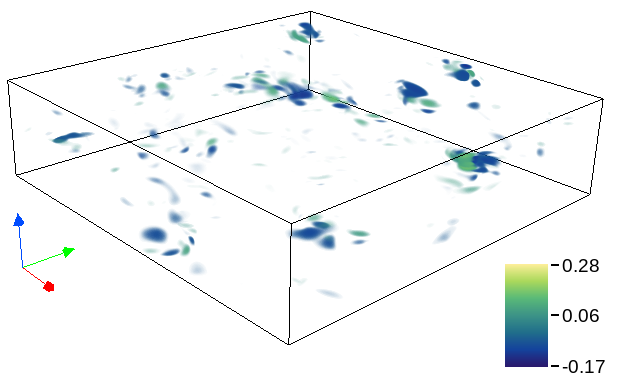}
	\put(-185,115){{\large{(a)}}}
	\includegraphics[width=0.4\linewidth]{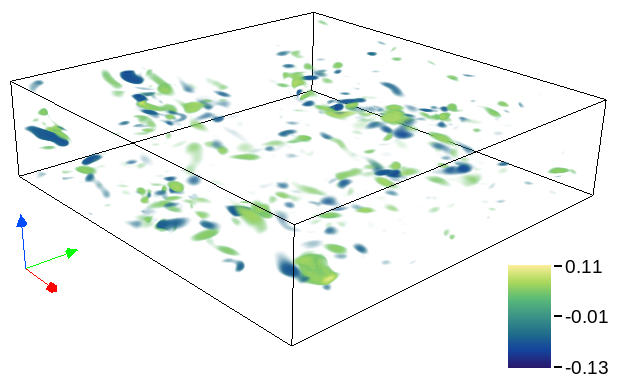}
	\put(-185,115){{\large{(b)}}}
	\caption{Extreme vertical velocity events under strong stratification.
		Pseudocolor plots of the vertical velocity field $w(\mbfx)$ at the maxima (a) and minima (b) of the kurtosis $K_w$. Only regions with $|w|\gtrsim 3\sigma$ are shown to highlight the most intense structures, where $\sigma$ denotes the standard deviation of $w$.
		}
	\label{fig:wextreme}
\end{figure}

These extreme events with very large kurtosis are associated with regions of intense updrafts and downdrafts. To illustrate such regions in the vertical velocity field for $\Fr=0.004$, we consider an instantaneous snapshot corresponding to a kurtosis extremum and highlight regions where $|w| \gtrsim 3 \sigma$. Figures~\ref{fig:wextreme}(a) and (b) show the vertical flow activity corresponding to the maximum and minimum of $K_w$, respectively. Snapshots with higher $K_w$ exhibit localized, intense regions of vertical motion, whereas those with smaller $K_w$ display a nearly homogeneous distribution of relatively weaker, spatially dispersed events. The presence of these spatio-temporal events makes the PDFs heavy-tailed, reflecting the strong intermittency of the vertical motions at small $\Fr$. Non-Gaussian PDFs are also observed for the density field, indicating that intermittency extends to scalar fluctuations as well (data not shown).

\begin{figure}
	\includegraphics[width=0.45\linewidth]{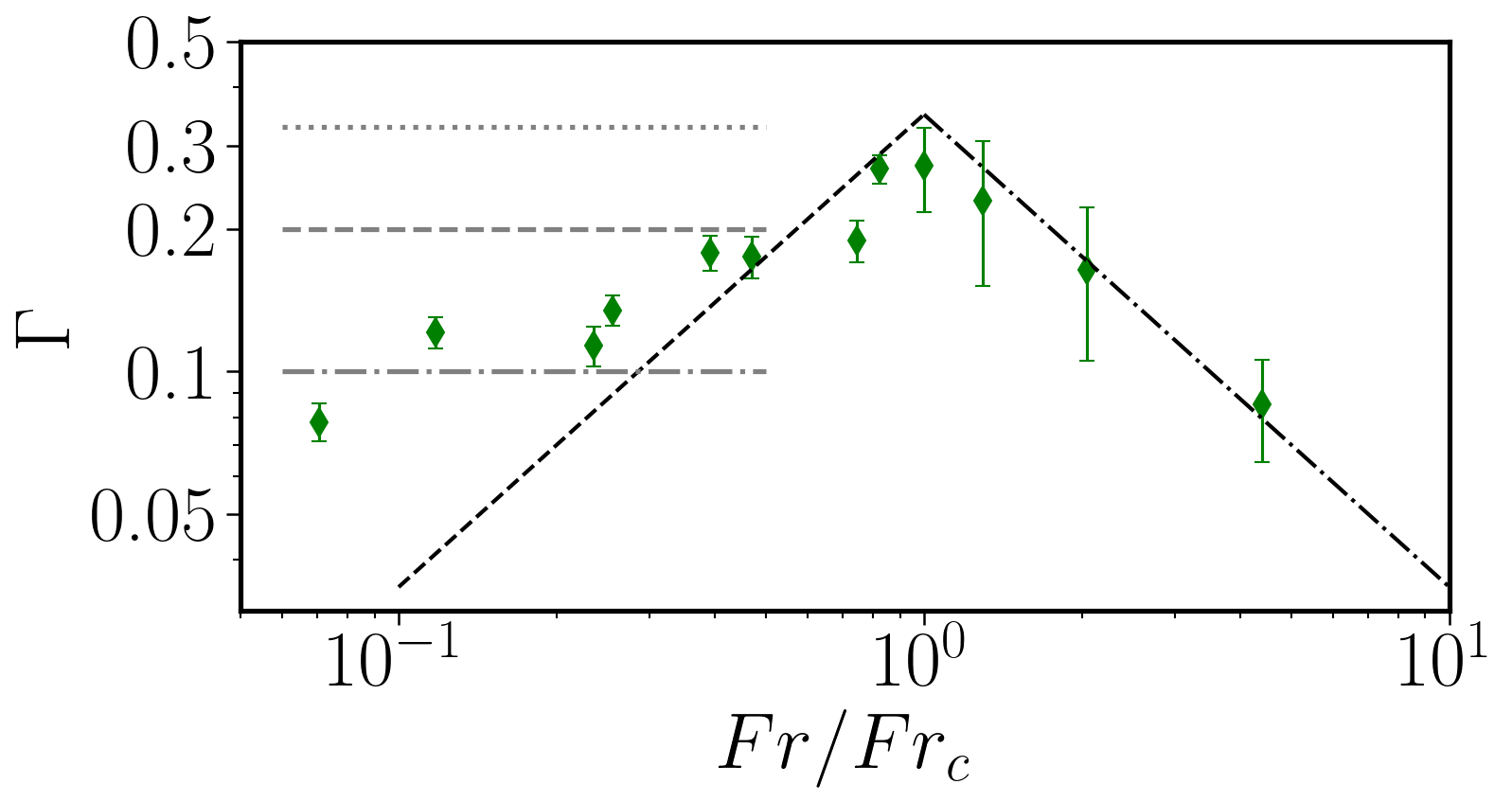}
	\put(-170,105){{\large{(a)}}}
	\includegraphics[width=0.45\linewidth]{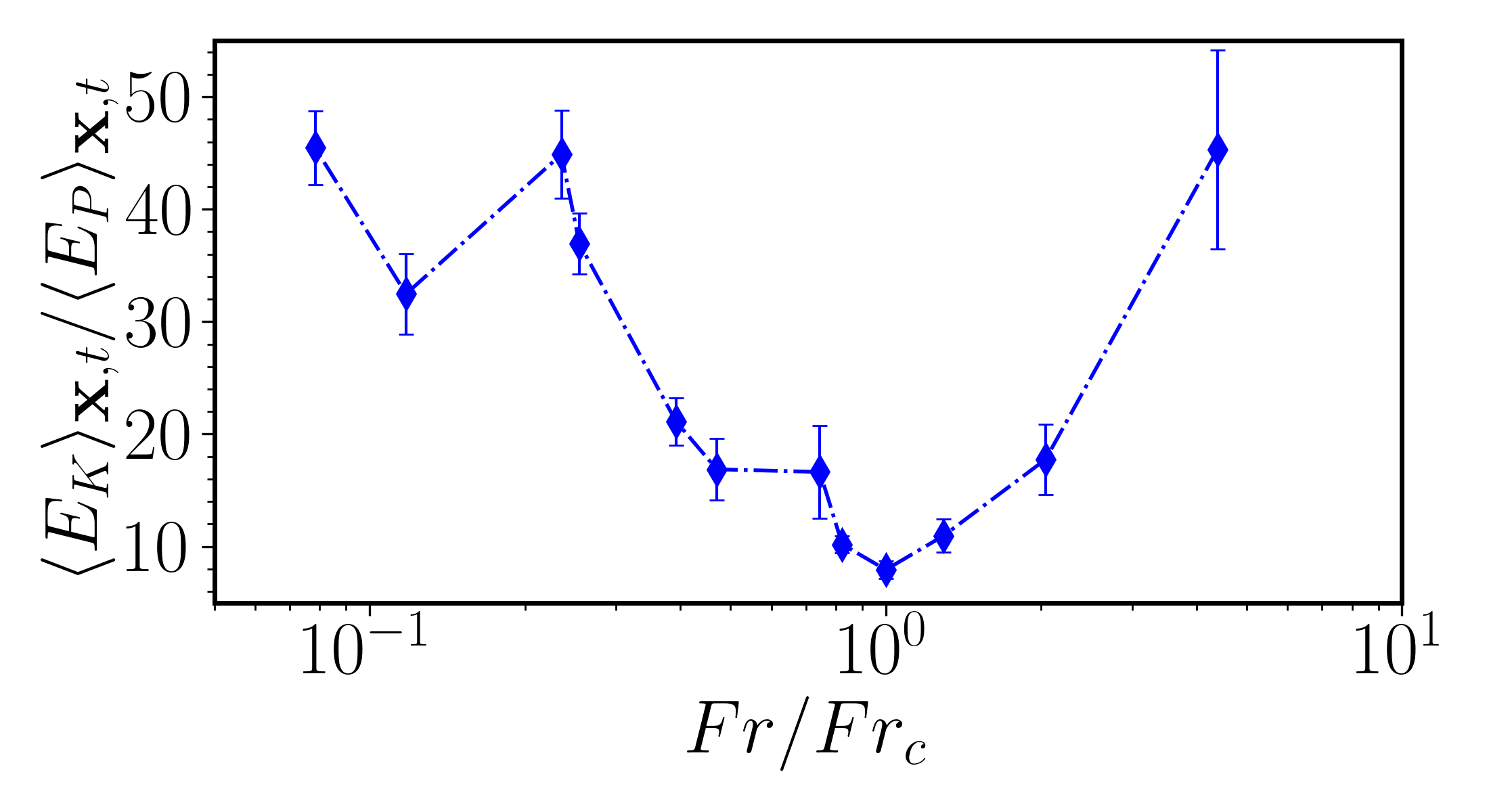}
	\put(-170,105){{\large{(b)}}}
	\caption{Mixing in stratified flows.
		(a) Mixing coefficient $\Gamma$ versus $\Fr/\Fr_c$. Horizontal lines indicate small-$\Fr$ limits reported in earlier studies: $\Gamma=0.1$ (dash–dotted line; Ref.~\cite{feraco2018vertical}), $\Gamma=0.2$ (dashed line; Ref.~\cite{osborn1980estimates}), and $\Gamma=0.33$ (dotted line; Ref.~\cite{maffioli2016mixing}).
		(b) Spatio-temporally averaged ratio of kinetic to potential energy, $\langle E_K \rangle_{\mbfx,t} / \langle E_P \rangle_{\mbfx,t}$, as a function of $\Fr/\Fr_c$. The critical Froude number $\Fr_c=0.051$ marks the transition to a nearly isotropic turbulent regime. Vertical bars represent the uncertainty in the mean, quantified by the standard deviation of the time series.
		}
	\label{fig:mixingeff}
\end{figure}

To investigate mixing, we examine the variation of the ratio $\Gamma = \langle  \epsilon_{\mathrm{dis}}^P \rangle_{\mbfx,t}/ \langle \epsilon_{\mathrm{dis}}^K \rangle_{\mbfx,t}$ as a function of $\Fr/\Fr_c$, as shown in Figure~\ref{fig:mixingeff}(a), where $\epsilon_{\mathrm{dis}}^P$ and $\epsilon_{\mathrm{dis}}^K$ denote the diffusive dissipation rate of potential energy and the viscous dissipation rate of kinetic energy, respectively. We find that the mixing coefficient $\Gamma$ lies in the range $0.08$–$0.25$. As $\Fr$ decreases from large values, $\Gamma$ increases and reaches a maximum at $\Fr=\Fr_c$, as indicated by the approximately linear trend shown by the black dash-dotted line. For $\Fr<\Fr_c$, $\Gamma$ exhibits non-monotonic variation with an overall slow decay. In a narrow range around $\Fr/\Fr_c \approx 0.5$, the values of $\Gamma$ cluster around $0.2$, whereas for stronger stratification ($\Fr/\Fr_c \lesssim 0.3$) they cluster around $0.1$, as indicated by the gray dashed and dash-dotted lines in Figure~\ref{fig:mixingeff}(a). Thus, increased stratification below the critical Froude number $\Fr_c$ leads to a reduction in mixing efficiency.

This trend is in qualitative agreement with the results of~\cite{feraco2018vertical}. The attainment of maximum mixing efficiency at $\Fr_c$ can be attributed to the enhanced role of Kelvin–Helmholtz instabilities. The subsequent decrease in $\Gamma$ with increasing $\Fr$ arises from the reduced ability of buoyancy forces to influence the turbulent dynamics. In this regime, turbulence becomes increasingly isotropic; although it efficiently stirs the density field, the weakening of buoyancy effects leads to diminished irreversible mixing.

Figure~\ref{fig:mixingeff}(b) shows the ratio $\langle E_K \rangle_{\mbfx,t} / \langle E_P \rangle_{\mbfx,t}$ as a function of $\Fr/\Fr_c$, where the total potential energy $E_P$ includes both reversible and irreversible components. For $\Fr>\Fr_c$, this ratio decreases monotonically with decreasing $\Fr$ until $\Fr=\Fr_c$, where it reaches a minimum value of approximately $10$. For $\Fr<\Fr_c$, the ratio $\langle E_K \rangle_{\mbfx,t} / \langle E_P \rangle_{\mbfx,t}$ exhibits non-monotonic behaviour with an overall increasing trend, and appears to saturate at a value of approximately $40$ under strong stratification. We note that the values of $\langle E_K \rangle_{\mbfx,t} / \langle E_P \rangle_{\mbfx,t}$ reported here are obtained by averaging over the statistically steady state reached at late times in our simulations.

\subsection{Discussion and conclusion}

We perform direct numerical simulations with sheared Kolmogorov forcing at a fixed geometrical aspect ratio and nearly constant Reynolds number to investigate the flow regimes that emerge as the stratification strength is varied over a wide range. The aspect ratio considered here corresponds to a 3D flow in the absence of stratification. As $\Fr$ increases, the system transitions from a buoyancy-dominated, strongly stratified regime to an instability-prone regime characterized by Kelvin–Helmholtz instabilities and enhanced mixing, and finally to a nearly isotropic turbulent regime.

In previous studies, a variety of forcing mechanisms have been employed to drive turbulence, such as fully 3D random forcing~\cite{feraco2018vertical}, purely vortical forcing~\cite{lucas2017layer, maffioli2016mixing}, quasi-2D random forcing with a small 3D component~\cite{lindborg2006energy}, and shear forcing~\cite{lucas2017layer,yi2023underlying}. Many of these studies have reported a leakage of energy toward VSHFs, which often come to dominate the dynamics once formed. Moreover, we did not modulate the forcing amplitude to maintain a prescribed energy injection rate. Instead, the flow was evolved over long durations to ensure that the statistical observables correspond to quasi-stationary states across a broad range of $\Fr$. In the absence of controlled energy input the flow lacks sharp interfaces separating density layers at small $\Fr$, as intermittent turbulent bursts tend to erode or disrupt the layering~\cite{lucas2017layer}. Additionally, we retained the dissipation terms in their standard form and did not replace them with artificial or hyperviscous operators. Although this choice reduces the extent of the observed inertial range, it ensures that the dynamics remain free from non-physical suppression of small-scale motions, which can arise as an artefact of artificial dissipation. 

Our results show that the energy associated with vertical motions scales as $\Fr^2$ for $\Fr<\Fr_c$, whereas for $\Fr>\Fr_c$ it becomes nearly constant, approaching the value obtained in the absence of stratification. The behaviour for $\Fr<\Fr_c$ is consistent with the analysis reported in Ref.~\cite{brethouwer2007scaling} across different Reynolds numbers. We have also analysed the partitioning of energy among vortical, wave-like, and mean VSHF modes. Both the energy spectra and fluxes exhibit a strong dependence on stratification strength, as do intermittent vertical transport and mixing.

In particular, we observe steeper reduced 1D perpendicular spectra. Similar steep spectra were reported in Ref.~\cite{laval2003forced, waite2004stratified, praud2005decaying,kitamura2006kh}. Our results differ from those reported in Ref.~\cite{brethouwer2007scaling}, even for $R_b\gtrsim1$.

We also computed the gradient Richardson number to assess the local development of Kelvin–Helmholtz instabilities in the flow fields (not shown). Such instabilities are triggered when the gradient Richardson number falls below the critical value of $0.25$~\cite{peltier2003mixing}, leading to enhanced overturning and mixing. We further observe that the vertical velocity field exhibits strong intermittency, consistent with the findings of Ref.~\cite{feraco2018vertical}. The kurtosis of the vertical velocity fluctuations deviates significantly from the Gaussian value ($K_w=3$) and attains large values at small $\Fr$, increasing monotonically as $\Fr$ decreases. This behaviour reflects the prevalence of intermittent, localized vertical bursts under strong stratification. In contrast, Ref.~\cite{feraco2018vertical} reported that the kurtosis initially increased with decreasing $\Fr$, reached a maximum, and then decreased toward the Gaussian value.

For strongly stratified flows at small $\Fr$, diminished vertical transport reduces the efficiency with which turbulent kinetic energy is converted into irreversible potential energy. Consequently, in the limit of small $\Fr$, fluid layers become increasingly resistant to mixing, leading to a saturation of the mixing coefficient. As the stratification weakens with increasing $\Fr$ toward $\Fr_c$, shear instabilities, specifically Kelvin–Helmholtz instabilities, enhance overturning and mixing. To obtain an unambiguous measure of irreversible conversion, we therefore adopt an energetics-based definition of mixing based on the ratio of the dissipation rates of turbulent potential and kinetic energies~\cite{salehipour2015diapycnal,maffioli2016mixing}. 

In oceanography, the diapycnal turbulent eddy diffusivity is often taken to be proportional to the ratio of the turbulent kinetic energy dissipation to the square of the local buoyancy frequency, with the proportionality constant serving as the mixing coefficient, typically assumed to have a constant value of $0.2$~\cite{osborn1980estimates}. However, this approach does not accurately separate the underlying energy exchange processes.
Ref.~\cite{maffioli2016mixing} showed that the mixing coefficient scales as $\Gamma \sim \Fr^{-2}$ and saturates at a value of $0.33$ in the small-$\Fr$ limit. In contrast, Ref.~\cite{feraco2018vertical} reported that $\Gamma$ varies linearly with the Froude number, $\Gamma \sim \Fr$, and approaches a saturation value of about $0.1$ for small $\Fr$. Our results are consistent with those of Ref.~\cite{feraco2018vertical} in the small-$\Fr$ limit, where the saturation constant is estimated to be approximately $0.1$. However, although $\Gamma$ exhibits a tendency to flatten at small $\Fr$, the available data points are insufficient to confirm this conclusively.
For the range $0.2 < \Fr/\Fr_c < 1$, a roughly linear increase in $\Gamma$ may be assumed, consistent with the behaviour reported in Ref.~\cite{feraco2018vertical}, although this trend remains far from conclusive. Moreover, both the peak value at $\Fr_c$ and the subsequent decline differ substantially from that study. We note that $\Gamma$ depends on the flow regime, control parameters, and the mechanism of turbulence generation~\cite{yi2023underlying}. Therefore, the observed mismatch could arise from several factors, such as differences in the forcing mechanism or the presence of VSHFs.

We note that exploring a wider parameter space ($\mathrm{Pr}, \, \Rey, \, \Fr$) is important for assessing the generality of our results. However, computational studies such as ours are often constrained by available resources, limiting access to regimes directly comparable to those encountered in geophysical flows. The extent of the parameter range over which our findings remain valid, together with the corresponding spectral behaviour in the presence of rotation superimposed on stratification under sheared forcing, remains to be explored.

\subsection*{Acknowlegments} Authors acknowledge National Supercomputing Mission (NSM) for providing computing resources of ‘PARAM Shakti’ at IIT Kharagpur and ‘PARAM Porul’ at NIT Trichy, which is implemented by C-DAC and supported by the Ministry of Electronics and Information Technology (MeitY) and Department of Science and Technology (DST).

\bibliographystyle{apsrev4-1}%
\bibliography{reference}

\end{document}